\documentclass[10pt, conference, letterpaper]{IEEEtran}

\usepackage{cite}
\usepackage{amsmath,amssymb,amsfonts}
\usepackage{algorithmic}
\usepackage{graphicx}
\usepackage{textcomp}
\usepackage{xcolor}
\def\BibTeX{{\rm B\kern-.05em{\sc i\kern-.025em b}\kern-.08em
    T\kern-.1667em\lower.7ex\hbox{E}\kern-.125emX}}
\usepackage{xspace}
\usepackage{setspace}
\usepackage{booktabs}
\usepackage{makecell}
\usepackage{cite}
\usepackage{amsmath,amssymb,amsfonts}
\usepackage{algorithmic}
\usepackage{graphicx}
\usepackage{textcomp}
\usepackage{subcaption}

\usepackage{xcolor}
\def\BibTeX{{\rm B\kern-.05em{\sc i\kern-.025em b}\kern-.08em
    T\kern-.1667em\lower.7ex\hbox{E}\kern-.125emX}}
\usepackage{wrapfig}
\usepackage{makecell}
\usepackage{comment}
\usepackage{blindtext}
\usepackage{graphicx}
\usepackage{booktabs} % For formal tables
\usepackage{amsfonts}
\usepackage{amsmath} 
\usepackage{booktabs}
\usepackage{multirow}
\usepackage{graphicx}
\usepackage{graphics}
\usepackage{xcolor}
\usepackage{color}
\usepackage{hyperref}
\usepackage{booktabs} % For formal tables
\usepackage{enumitem}
\usepackage{xspace}
\usepackage{balance}
\usepackage{blindtext}
\renewcommand{\baselinestretch}{0.97}
\usepackage{cite}
\usepackage{amssymb}
\usepackage{algorithmic}
\usepackage{textcomp}
\usepackage{pifont}
\newcommand{\xmark}{\ding{55}}%

\newcommand{\etal}{{\em et al.}\xspace}
\newcommand{\eg}{e.g.,\xspace}
\newcommand{\ie}{{\em i.e.,}\xspace}

\newcommand{\BfPara}[1]{\vspace{0.2em}{\noindent\bf#1.}\xspace\xspace}

\usepackage{tikz}
\DeclareMathAlphabet{\mathpzc}{OT1}{pzc}{m}{it}

\usepackage{fontenc}
\usepackage{amssymb}

\usetikzlibrary{shapes.misc}

\begin{document}
\title{ML-based IoT Malware Detection Under Adversarial Settings: A Systematic Evaluation}
% \title{Systematically Evaluating the Robustness of ML-based IoT Malware Detection Systems}

\author{\IEEEauthorblockN{Ahmed Abusnaina$^{\dagger}$, Afsah Anwar$^{\dagger}$, Sultan Alshamrani$^{\dagger}$, Abdulrahman Alabduljabbar$^{\dagger}$,\\ Rhongho Jang$^{\diamond}$, Daehun Nyang$^\ddagger$, and David Mohaisen$^\dagger$}\\
\IEEEauthorblockA{$^\dagger$University of Central Florida, $^{\diamond}$ Wayne State University, $^\ddagger$ Ewha Womans University \\ \textit{ahmed.abusnaina, afsahanwar, salshamrani, jabbar @knights.ucf.edu; r.jang@wayne.edu;} \\ \textit{nyang@ewha.ac.kr; mohaisen@ucf.edu} }
}

% \author{ 
% \IEEEauthorblockN{Hisham Alasmary}
% \IEEEauthorblockA{
% \textit{University of Central Florida}\\
% Orlando, Florida, US \\
% hisham@knights.ucf.edu}
% \and
% \IEEEauthorblockN{Mohammed Abuhamad}
% \IEEEauthorblockA{
% \textit{University of Central Florida}\\
% Orlando, Florida, US \\
% abuhamad@knights.ucf.edu}
% \and
% \IEEEauthorblockN{Ahmed Abusnaina}
% \IEEEauthorblockA{
% \textit{University of Central Florida}\\
% Orlando, Florida, US \\
% ahmed.abusnaina@knights.ucf.edu}
% \and
% \IEEEauthorblockN{Afsah Anwar}
% \IEEEauthorblockA{
% \textit{University of Central Florida}\\
% Orlando, Florida, US \\
% afsahanwar@knights.ucf.edu}
% \and
% \IEEEauthorblockN{Aziz Mohaisen}
% \IEEEauthorblockA{
% \textit{University of Central Florida}\\
% Orlando, Florida, US \\
% mohaisen@ucf.edu}
% }

\maketitle

\begin{abstract}
The rapid growth of the Internet of Things (IoT) devices is paralleled by them being on the front-line of malicious attacks. This has led to an explosion in the number of IoT malware, with continued mutations, evolution, and sophistication. These malicious software are detected using machine learning (ML) algorithms alongside the traditional signature-based methods. Although ML-based detectors improve the detection performance, they are susceptible to malware evolution and sophistication, making them limited to the patterns that they have been trained upon. This continuous trend motivates large body of literature on malware analysis and detection research, with many systems emerging constantly, outperforming their predecessors.

In this work, we systematically examine the state-of-the-art malware detection approaches, that utilize various representation and learning techniques, under a range of adversarial settings. Our analyses highlight the instability of the proposed detectors in learning patterns that distinguish the benign from the malicious software. The results exhibit that software mutations with functionality-preserving operations, such as stripping and padding, significantly deteriorate the accuracy of such detectors. Additionally, our analysis of the industry-standard malware detectors shows their instability to the malware mutations.

Through extensive experiments, we highlight the gap between the capabilities of the adversary and that of the existing malware detectors. The evaluations and analyses show that the optimal malware detection system is nowhere near, and calls for the community to streamline their efforts towards testing the robustness of malware detectors to different manipulation techniques.

\end{abstract}


% \vspace{-3mm}
\section{Introduction}

IoT malware have been the focus of the security research community and the industry alike. 
These efforts have resulted in various malware detection approaches, intended for safeguarding the IoT infrastructure against increasing targeted attacks. These proposed detectors leverage the traditional signature-based approach or the capabilities of the learning algorithms to build Artificial Intelligent (AI)-based detectors. These detection systems leverage modalities generated through static and dynamic software analysis techniques, along with deep learning and natural language processing, for generalizing detection to previously unseen IoT malware~\cite{PajouhDKC18}. 
% Those engines that feed into the likes of VirusTotal are fittingly considered as the up-to-date capability of industry-standard malware detectors.

Considering that these techniques are heavily dependent on the specific data used for their training and testing, it is plausible that they would have a reduced performance when tested in an uncontrolled environment due to various practical settings. For example, the constant evolution of malware that employ obfuscation may impact the performance of these detectors over time, especially the static-based techniques. 
% Moreover, packed software samples are known to be categorized as malicious by the industry-standard malware detectors~\cite{aghakhani2020malware}. 
While packing is widely used among malicious software, it is not exclusive to malware. This limit the usage of packing as a detection modality, since that may result in a large number of false positives. 
Even in the absence of packing, malware detection systems have been shown to be susceptible to adversarial attacks. An adversary can manipulate the features of any software, directly or indirectly, to force the detector to output the adversary's desired decision~\cite{severi2020exploring,AbusnainaKAPAM19,KreukBABPK18}.

A common practice for inspecting software is using online scan engines, such as VirusTotal~\cite{VirusTotal}, which embody the aforementioned techniques for malware detection and provide reports that contain the detection results of a pool of anti-virus engines. Additionally, these online scanners are utilized by the malware developers to check if their malicious payloads can evade detection from the anti-virus engines before starting a malware campaign~\cite{GrazianoCBL15}.
Altogether, before deploying such malware detection systems in practice, it is essential to understand the shortcomings of state-of-the-art IoT malware detection systems under adversarial settings that can be abused by the adversaries towards future malware campaigns.

In this work, we examine state-of-the-art malware detection approaches, including those that rely on different representation and learning algorithms. We consider techniques that represent the software as a binary sequence, static disassembly features, and graphs.
These representations yield a promising detection performance, with higher than 99\% detection accuracy~\cite{mercaldo2020deep,LiSYLSY18,AhmadiUSTG16,xu2018cdgdroid,ChenYHWPWY18}. However, our findings highlight the instability of the learning algorithms in learning useful fundamental patterns that represent the difference between benign and malicious software (more details can be found in~\autoref{sec:results}). 

\begin{figure}[t]
    \centering
    \includegraphics[width=0.45\textwidth]{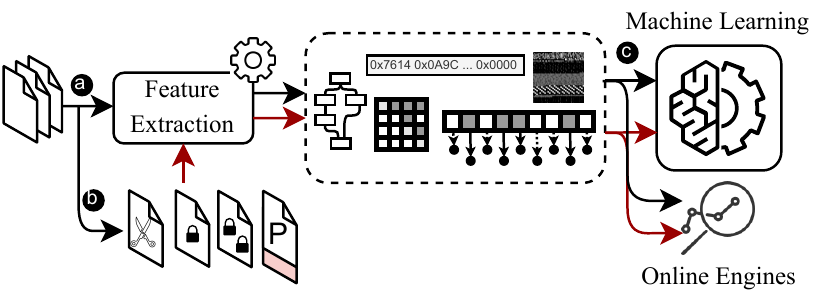}\vspace{-1mm}
    \caption{The system pipeline. The software binaries are (a) represented using different state-of-the-art approaches, and (b) manipulated using functionality preserving operations, such as packing, stripping, and padding. The corresponding representations of the original samples and manipulated ones are then (c) tested against pre-trained ML-based malware detectors and industry-standard detection engines.}
    \label{fig:pipeline}\vspace{-3mm}
\end{figure}

By systematically evaluating the robustness of various malware detectors, we demonstrate that manipulating the malicious software with functionality-preserving operations, such as stripping and binary padding, significantly reduces the detectors' performance. Towards this, we generate four equivalent binaries for each software using means of packing (with different compression levels), stripping, and padding. 
% The generated binaries are executable, and preserve the functionality of the original software. 
We evaluate each of the resultant software against various IoT malware detection approaches, along with the industry-standard malware detection engines. 
The results show a concerning behavior, where one or more detectors fail to hold a reasonable performance (lower than 50\% detection rate) in detecting malware mutations.
~\autoref{fig:pipeline} shows the different phases of analysis strategy; feature representation, software manipulation, and evaluation of ML-based malware detectors.

\BfPara{Contributions} This work highlights the discrepancies between the capabilities of the adversary and the assumed adversarial capabilities by the research community. Particularly, we make the following contributions:
\begin{enumerate}%[leftmargin=*]
    \item \underline{\textit{Validity of the baseline}}: We examine nine state-of-the-art malware detection representations and three learning algorithms and evaluate their performance using a total of 5,295 IoT software binaries. The evaluation shows the effectiveness of each representation in detecting malicious IoT software with high accuracy in a level playing field.
    \item \underline{\textit{Model instability}}: We investigate the stability of the baseline malware detectors. Our results demonstrate the inconsistency of the learning process, \ie with the introduction of a small random perturbation to the input space, the detector is rendered useless (outputs random label).
    \item \underline{\textit{Vulnerability to adversarial settings}}: We examine the robustness of the IoT malware detectors under white-box and black-box adversarial settings, resulting in an accuracy reduction of up to 70\%.
    \item \underline{\textit{Vulnerability to binary manipulation}}: We evaluate detectors against three  manipulation techniques: packing, stripping, and padding. These techniques are practicality and functionality preserving, where the generated software is identical in functionality to the original software. Our evaluation shows that such software is capable of misleading the state-of-the-art malware detectors.
    \item \underline{\textit{Vulnerability of industry-standard malware detectors}}: The evaluation of industry-standard malware detection engines shows that most of the engines are rendered useless upon slight modification of the software.
\end{enumerate}

% \BfPara{Organization} The rest of this paper is organized as follows. We provide a background on malware detection and evasion techniques in~\autoref{sec:background}. We discuss the threat model under which we evaluate the robustness of malware detectors in~\autoref{sec:threat}. Overview of the used dataset is provided in~\autoref{sec:Dataset}. Then, we evaluate the state-of-the-art malware detectors (\autoref{sec:results}) and the industry-standard detection engines (\autoref{sec:onlineengines}). We conclude our work in~\autoref{sec:conclusion}, providing the main takeaways of this study.

\section{Background}\label{sec:background}

The increasing security concern for IoT devices has been paralleled by an increasing body of work in the area of IoT security, particularly addressing malware analysis and detection. Building towards our work, it is important to outline the efforts that propose IoT malware detection systems and the methods of evasion that will elucidate the susceptibility of the malware detection systems to various adversaries.
% In this section, we revisit some of those efforts that propose IoT malware detection systems.

\subsection{Malware Detection}\label{sec:malware_detection_background}
Prior works have shown the potential and feasibility of ML to detect malware with more than 99\% accuracy~\cite{AlasmaryAPCNM19,WangCYJPYC20,AnwarAPW20,YanCZPYHZY21,vasan2020image,mercaldo2020deep,LiSYST19}. The performance of these detection systems depends on the choice of software representations, which are a result of two common analysis techniques. 
In \textit{dynamic analysis}, a malware is executed in a monitored sandbox environment. The behavioral patterns are then used as feature representation. However, dynamic analysis is time and space-consuming, thereby limiting its scalability~\cite{willems2007toward}.

The \textit{static analysis} involves analyzing the binary executable without executing it. The fast and scalable extraction of representations makes static analysis the primary analysis technique for malware detection. Malware binaries have multiple features that can be statically extracted and used as modalities for malware representation.

\BfPara{Selected Representations}
We focus on representations that are (1) extensively used in the prior works, (2) fast to generate, and (3) can be extracted for malware detection on the fly. We summarize the used representations in the following.
    \begin{enumerate}[topsep=0.5pt,leftmargin=*]
        \item A common strategy is to transform the malware into a \textit{grayscale image}. Particularly, the byte-code is visualized as a grayscale image of a fixed size of ($h\times w$) where every byte is a pixel in the image.
        \item \textit{CFG adjacency}. Another strategy is to extract the assembly instructions by disassembling malware and further transforming them into a Control Flow Graph (CFG) by dissecting them into basic blocks depending on the instruction branching or jumps. The CFG is then represented as a square matrix representing edges between nodes.
        \item \textit{CFG algorithm}. Graph algorithms have been augmented to extract graph attributes that represent the connectivity patterns in the CFG. 
        % These features are exhibited in~\autoref{tab:DSFeatures}. 
        \item \textit{Strings} are a sequence of printable characters in the binary codebase. The strings of a program are analyzed to understand the possible behavioral patterns of the malware and can also be used to prepare a sandbox environment for the dynamic analysis~\cite{CozziGFB2018}.
        \item \textit{Segments} are necessary for program execution. They describe the memory layout of an executable and is interpreted by the kernel during load~\cite{ONeill16}. Within every segment, there may be code or data divided among \textit{sections}, such as \textit{.text}. Binaries contain symbol tables which are used as references for linking and debugging~\cite{ONeill16}.
        \item \textit{Symbols} are symbolic references to code or data and include global variables or functions. Every executable generally has two symbol tables: the symbol table that contains all symbol references and the dynamic symbol table which only contains references for dynamic symbols~\cite{ONeill16}.
        \item \textit{Hexdump} represents a malware as a sequence of hexadecimal values, where each value represents two bytes (in 0-255 range), the frequency of which is then recorded as a vector of size $1\times 256$. 
        \item \textit{Feature fusion} represents a unified (combined) representation of all of the aforementioned representations.   
    \end{enumerate}

For the completeness of the study, we include malware representations proposed by works that are not strictly IoT malware-specific. ~\autoref{tab:RepresentationsLet} summarizes the malware representations that have been proposed for malware detection, and utilized in this work.

% \begin{table}[t]
%     \centering
%     \caption{The CFG extracted algorithmic features, categorized into seven groups. When possible, the minimum, maximum, median, mean, and standard deviation are calculated.}\vspace{-1mm}
%     \label{tab:DSFeatures}
%     \scalebox{1.0}{

%     \begin{tabular}{|l|c|}
%         \Xhline{3\arrayrulewidth}
%         Feature category & \# of features\\
%         \Xhline{2\arrayrulewidth}
%         Betweenness centrality &  5  \\ \Xhline{1\arrayrulewidth}
%         Closeness centrality &  5  \\  \Xhline{1\arrayrulewidth}
%         Degree centrality &  5  \\  \Xhline{1\arrayrulewidth}
%         Shortest path &  5  \\  \Xhline{1\arrayrulewidth}
%         Density &  1  \\  \Xhline{1\arrayrulewidth}
%         \# of Edges &  1  \\  \Xhline{1\arrayrulewidth}
%         \# of Nodes &  1  \\  \Xhline{2\arrayrulewidth}
%         Total & 23 \\ \Xhline{3\arrayrulewidth}
%     \end{tabular}}\vspace{-2mm}
% \end{table}

\subsection{Representation Evasion}\label{sec:representation_evasion_background}
Several software evasion and manipulation techniques were proposed for malware mutation and misclassification. In the following, we briefly discuss the commonly used   techniques.

\BfPara{Binary Packing}
Packing is used by malware authors to thwart detection or analysis by detectors, analysts. The packer is augmented to compress or encrypt an executable, where the code and data are hidden from the analysts. Considering that portions of the executable are compressed, it needs to be decompressed before it is executed in memory~\cite{ONeill16}.

Typical packing software consist of two programs, packer program and the stub program, where the first packs the software while the second deobfuscates the software. While there are many packing programs, such as \textit{DacryFile} by Grugq, \textit{Burneye} by Scut, \textit{Shiva} by Neil and Shawn, and \textit{Maya's Veil} by Ryan, the \textit{Ultimate Packer for eXecutables} (UPX)~\cite{upx} is the one most commonly used~\cite{CozziGFB2018}. UPX utilizes the UCL data compression library algorithm~\cite{uclCompression} which uses in-place decompression, and does not introduce memory overheads.

\BfPara{Binary Stripping}
Stripping is utilized to hide  information that may leak the functional software strategies. A codebase can be compiled with no standard library linking (\textit{gcc-nostlib}). Alternatively, parts of the ELF file can be hidden such that the different constituents of the binary format can be obfuscated such that the interpretation can be halted. The resultant binaries would be void of information such as debug and relocation information, section headers, and symbols~\cite{stripping}.

\begin{table}
\centering
\caption{The state-of-the-art static analysis representations used in this work. Most of the representations require reverse-engineering (R.E.), while image-based representation directly used the raw binaries (Bin.). CODE: features extracted from the disassemble binaries.}\vspace{-1mm}
\label{tab:RepresentationsLet}
\scalebox{0.85}{
\begin{tabular}{|l|l|c|c|c|c|}
% \hline
% \hline
\Xhline{2\arrayrulewidth}

Type & Feature & Work & Bin. & R.E. & Graph \\
% \hline
\Xhline{2\arrayrulewidth}
Binary & Image    &   \cite{kancherla2013image,vasan2020image,yajamanam2018deep,mercaldo2020deep}  &  $\checkmark$   & \xmark & \xmark \\
\hline
CFG & Adjacency    &  \cite{xu2018cdgdroid,jalote2012integrated,bruschi2006detecting}  &  \xmark & $\checkmark$  &  $\checkmark$   \\\hline
CFG & Algorithmic     & \cite{AlasmaryAPCNM19,bruschi2006detecting,AnwarAPW20}   & \xmark  & $\checkmark$   &  $\checkmark$  \\\hline
CODE & String     &  \cite{AhmadiUSTG16,AnwarAPW20}  &   \xmark  & $\checkmark$ & \xmark \\\hline
CODE & Symbols    &   \cite{AhmadiUSTG16,AnwarAPW20} &   \xmark  & $\checkmark$  & \xmark \\\hline
CODE & Sections    & \cite{AhmadiUSTG16,AnwarAPW20}   &  \xmark   & $\checkmark$  & \xmark \\\hline
CODE & Segments    &  \cite{AhmadiUSTG16}  &   \xmark  & $\checkmark$ & \xmark \\\hline
CODE & Hexdumps     &  \cite{AhmadiUSTG16}  &  $\checkmark$   & $\checkmark$ & \xmark \\\hline
CODE & Combined     & \cite{AhmadiUSTG16,AnwarAPW20}   &  $\checkmark$   & $\checkmark$ & \xmark \\
\Xhline{2\arrayrulewidth}
\end{tabular}
}\vspace{-2mm}
\end{table}

\BfPara{Adversarial Evasion}
With the growth in ML adoption, it is essential to understand and assess the robustness of ML techniques to several adversarial settings.
These settings include adversarial examples, in which an adversary crafts perturbation to misguide the model output to its desired label by applying a minimal perturbation to the original sample~\cite{PapernotMJFCS16}.  

Given a model objective function $f(.)$ and a sample represented by a vector $x$, the adversary aims to introduce perturbation ($\delta$) in the feature space $x' = x + \delta$ such as $f(x) \neq f(x')$. Crafting the perturbation can be derived from two perspectives: targeted and non-targeted attacks. 
\BfPara{Targeted attacks} The adversary in this attack generates an adversarial example $x'$ that forces the classifier to misclassify into a specific target class $t$. For instance, the adversary generates a set of malicious IoT software samples, which are classified as benign. That is: $x': [f\left ( x' \right) = t]$.
\BfPara{Untargeted attacks} The adversary's goal is to misclassify the output of the model to any class other than the original label. That is  $x': [f\left ( x' \right) \neq f\left ( x \right) ]$. In this work, we only consider the two-class classification task, where targeted and untargeted attacks behave the same.

Adversarial attacks can be launched under different adversarial capabilities that allow for either black-box or white-box attacks. In a white-box attack, the adversary has full knowledge of the inner networking paradigm of the model. In a black-box attack, the adversary has only access to the model via an oracle and can only observe the model's output. 

Several methods have been proposed to generate adversarial examples by directly perturbing the feature space in both black-box and white-box settings~\cite{GoodfellowSS15,HuT17,Moosavi-Dezfooli16,KurakinGB17a}. For example, Carlini and Wagner~\cite{CarliniW17} proposed generic adversarial attacks against distilled Neural Networks (NN), which showed its effectiveness against several ``robust'' deep learning NN. 

While initially designed to exploit image-based classifiers, where perturbation can be directly applied to the image pixels~\cite{PapernotMJFCS16,PapernotMGJCS17,wangYVZZ18}, adversarial attacks showed high success in malware detection while preserving the software functionality and executability~\cite{grosse2017adversarial,AbusnainaKAPAM19}. At the binary-level, several studies~\cite{KolosnjajiDBMGE18,KreukBABPK18} generated practical adversarial examples by appending binaries to the original file. While it is effective against signature- and binary-based classifiers, it can be countered by reverse-engineering the software to extract the corresponding representations. 

Other studies~\cite{AbusnainaAASNM19,AbusnainaKAPAM19} introduced adversarial attacks on the execution flow of the code, by injecting benign functionalities within the malware and vice versa. However, such a perturbation should be applied to the source code, and is only possible by the malware author, unlike the binary padding approach. 

To investigate the effectiveness of different malware representation and learning approaches, we examine a wide set of adversarial settings, including direct generic and modified adversarial attacks, as well as the black-box adversarial settings.

\section{Threat Model}\label{sec:threat}
Learning algorithms are widely used to obtain state-of-the-art performance in several fields, including malware detection. However, the usage of ML in critical domains is subject to adversarial attacks. 
% In this work, we consider using white-box configurations whenever possible, while evaluating the systems against different black-box attacks. 
In the following, we discuss the threat models used for systematically evaluating the robustness of the malware detectors.

\subsection{Gaussian Noise}
A stable learning model is argued to be immune to misclassification under the introduction of Gaussian noise in the feature space, as unguided perturbation is unlikely to disrupt the existing patterns to some extent~\cite{hu2019new,guo2018countering,SzegedyZSBEGF13}. 

A correctly trained model that can distinguish benign and malicious samples with high confidence, is constraint by three factors. (1) \textit{Data representation}: A robust software representation should contain meaningful patterns that can distinguish the malicious from the benign software, (2) \textit{Learning algorithm}: The learning algorithm should be able to capture such patterns even at a higher dimensionality without over-fitting or under-fitting, and (3)  \textit{Training data}: The trained model should be generalizable to unseen new samples, and samples that are not fundamentally different from the ones in the training dataset.
This requires the training data to be cohesive and the samples of each class to be an accurate representation of that class. While the first two factors are considered, the third is an open challenge, and we consider it out-of-scope of this work.

In this work, we use the Gaussian noise as a metric to measure the stability of the representations. Given the model objective function $f(.)$, data points (samples) $x \in X$ with feature space of $n$ features, the output of the model is defined as $y = f(x)$. The Gaussian noise is then calculated as follows:
\[x_{i}' = x_{i} + max(X_{i}) \times \delta,\quad \forall i \in n,\]
where $X_{i}$ is a list of the $i^{th}$ features of all $x \in X$. A stable model is then defined as:
\[f(x) = f(x'),~~ \text{if} ~ \delta < threshold.\]
In this work, we do not introduce a cut-off threshold for a stable model. However, we observe the model's behavior when a perturbation in the range of [1\%, 100\%] is introduced. Ideally, the relationship between the accuracy and perturbation should be linear: with an increasing perturbation, the accuracy should linearly decrease, \eg to reach random (50\%) at 100\% perturbation given the two-class classification task. We note that this attack will not generate practical adversarial examples, as it applies the perturbation to the feature space directly. Rather, it is used to measure the detectors' stability.

\begin{figure}[t]
    \centering
    \includegraphics[width=0.45\textwidth]{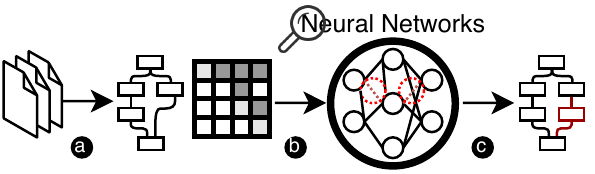}%\vspace{-3mm}
    \caption{Graph manipulation. The software is reverse-engineered and (a) represented as CFG and  adjacency matrix, (b) using the pre-trained neural network, (c) white-box C\&W-based perturbation is crafted/applied to the CFG. % We limit the allowed actions to adding edge and adding new node to generate a realistic CFG.
    }
    \label{fig:graphAdjAttack}\vspace{-5mm}
\end{figure}

\subsection{Graph Manipulation}
This configuration targets the graph-based representations, including the adjacency- and algorithmic-based representations extracted from the software's corresponding CFGs. 
Given a CFG $G = \{V,E\}$, where $V$ is the set of nodes in the graph, and $E$ is the set of edges, the adversary's goal is to introduce a carefully crafted perturbation that misclassifies the system to the desired output. To introduce such a perturbation, we used the adjacency matrix representation as a baseline to craft the perturbation. Then, the Carlini \& Wagner $L_{\infty}$ (C\&W) attack~\cite{Carlini017} is used to craft the perturbation under the white-box settings. The C\&W is a gradient-based attack that optimizes the penalty and distance metrics on $L_\infty$ norms in the process of generating adversarial examples. This method ensures that the added perturbation will be minimal while causing misclassification.

Using the adjacency matrix representation, the adversary aims to craft a perturbation $\delta \in \mathbb{R}^{d\times d}$ as a domain-specific range of possible features that can be observed in ordinary samples. 
This perturbation achieves the adversarial goal if 
$y = f(x) \neq f(x+\delta)$, 
where $y'$ is the classifier's prediction after applying the perturbation $\delta$ to the original feature space $x$.~\autoref{fig:graphAdjAttack} shows the outline of the attack. To keep the generated CFG realistic, we limit the actions done by C\&W attack to only adding nodes and edges. This is done by modifying the original attack to prevent deleting existing edges, and only limiting the process to adding edges.

While CFG manipulation preserves the original functionality~\cite{AbusnainaAASNM19,AbusnainaKAPAM19}, we do not have access to the source code of the samples. Therefore, we cannot generate practical adversarial binaries using CFG manipulation. Given that, we used this attack to evade the graph-based detectors using direct white-box attacks on NN-based adjacency matrix-based classifier, while transferring the attack to remaining CFG-based classifiers.

\subsection{Static String Manipulation}
Another white-box attack is the string manipulation attack. In this representation, the software is represented as a vector $V$ of bag of words $W$ of size $1\times |W|$, where $|W|$ is the number of words considered in the representation. Similar to the graph manipulation attack, we used C\&W $L_{\infty}$ attack to craft a minimal perturbation to misclassify the model. Given that the crafted perturbation cannot be applied directly to the binaries, we consider it as a practical attack under the assumption of the availability of the source code. We evaluate this attack by crafting the perturbation using the NN baseline and transferring the attack to the remaining baseline models.

\subsection{Binary Packing}
Recall that a binary executable can be packed using packer software, such as UPX (see \autoref{sec:representation_evasion_background}).
The ML-based detectors utilize the features, such as raw binaries, strings, and segments, from the malware. These features are, however, suppressed from packing. In this attack, we pack the malware and probe the performance of the representation used in the literature. Moreover, UPX supports different degrees of packing. For this study, we utilized the default settings and the best compression method of UPX.

\subsection{Binary Stripping}
Recall that a binary can be stripped of information without affecting its executability (see \autoref{sec:representation_evasion_background}). In this attack, we probe the impact of a stripped binary on an ML-based detector's performance. Particularly, we strip the binaries of their debug information and the symbol information that are not needed for relocation.

\begin{figure}[t]
    \centering
    \includegraphics[width=0.45\textwidth]{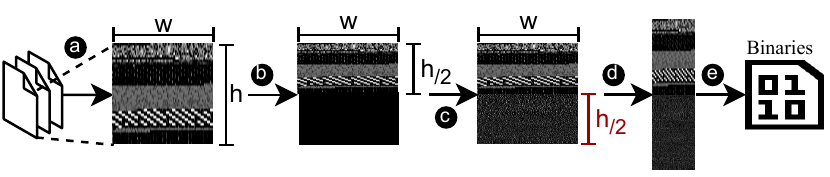}%\vspace{-3mm}
    \caption{Binary padding attack overview. (a) The software is represented as an $h\times w$ image. (b) The content of the image is then compressed into the size of $\frac{h}{2}\times w$. (c) Using C\&W attack, we generate perturbation on the remaining half $\frac{h}{2}\times w$ of the image. (d) The generated image perturbation is then rescaled to the original size of the software, and then (e) reshaped to a 1-D vector represented the binaries to be appended.}
    \label{fig:imageAttack}%\vspace{-2mm}
\end{figure}

\subsection{Binary Padding}
In this attack, the adversary aims to craft a white-box practical (executable) adversarial example by appending binaries to the end of the software binaries.~\autoref{fig:imageAttack} shows the process of generating perturbation in the white-box settings for image-based representation baselines. For a software $s$ of size $z_{s}$ represented as an image $img$ of size $h\times w$, we first compress the content of the image into the space $\frac{h}{2}\times w$. Afterward, we craft a minimal perturbation using C\&W attack. To prevent the attack from applying a perturbation to the upper half of the image, the attack is modified allowing changes in the lower half of the image. After the evasion, we convert the generated lower half of the image of size $\frac{h}{2}\times w$ back to the actual size $z_{s}$ of the software $s$, and then converting it to 1-D vector by concatenating the rows. We note that this attack will introduce a perturbation size of 100\%, as the perturbation has the same size as that of the original file, and the generated software $s'$ will be of size $z_{s'} = 2\times z_{s}$. This attack generates an adversarial software that is executable. We evaluate the generated software on the image-based baseline models, in addition to the other representations by re-extracting the features from the manipulated software.

\section{Dataset Overview}\label{sec:Dataset}
To analyze the robustness of state-of-the-art malware detectors, we start by collecting a dataset of malicious and benign IoT binaries. The dataset was collected between November 2018 and December 2020, where 3,000 malware samples of three families---Gafgyt, Mirai, and Tsunami---were retrieved from CyberIOCs~\cite{cyberiocs19}, VirusTotal~\cite{VirusTotal}, and VirusShare~\cite{VirusShare}, in addition to 2,295 benign samples, compiled from source files on GitHub~\cite{github19} with different optimization levels.

\BfPara{Ground Truth Class} We used {\em VirusTotal}~\cite{VirusTotal} to validate the malicious and benign samples in our dataset. The samples were first uploaded to VirusTotal. After 24 hours, the scan results corresponding to each sample were retrieved.

\BfPara{Data Augmentation} As aforementioned in~\autoref{sec:background}, the dataset samples are transformed to different representations: (1) Represented as images to be fed into an image-based classifier. (2) Using {\em Radare2}~\cite{radare2}, a reverse-engineering open-source framework for analyzing binaries, the samples were reverse-engineered to obtain various features, such as strings, symbols, sections, and segments.  (3) Hexdump representation is used to represent the ``\textit{.text}'' section of the binaries. (4) The software CFG is extracted using {\em Radare2}, which then used to generate the software adjacency matrix and different graph-theoretic features.
% these features are categorized under seven groups, where each group contains a category of features. The minimum, maximum, median, mean, and standard deviation values for the observed parameters were represented by the five features extracted from each feature category.

\section{Robustness Analysis}\label{sec:results}
%In the arm race between malware detectors and malware authors, malware detection and identification require an accurate understanding of the capabilities of malware authors. In this section, we evaluate the existing on-the-fly static-based malware detection techniques (see~\autoref{sec:malware_detection_background}) against executability- and functionality-preserving software binary manipulations.

\subsection{Experimental Setup}
Towards evaluating the robustness of the state-of-the-art IoT malware detectors, the dataset is transformed using the nine representations. Then, four learning algorithms are used to establish the baseline classifiers.

\BfPara{Learning Algorithms}
Several classification algorithms have been adopted and used in various domains in IoT malware detection and classification~\cite{AlasmaryAPCNM19,ShenHZYFC18}. 

In this study, we evaluate the robustness of four ML algorithms, namely, \textit{Logistic Regression (LR)}, \textit{Random Forest (RF)}, \textit{Convolutional Neural Networks (CNN)}, and \textit{Deep Neural Networks (DNN)}.
The selection of learning algorithms is for multiple reasons. They are (1) commonly used in this domain, (2) fundamentally different in the learning process, (3) highly sophisticated approaches, such as DNN and CNN, and simpler ML algorithms, such as LR and RF. For instance, the LR-based classifier is selected to extract the relationships between variables in the feature space, with no deep representations.  CNN, on the other hand, was selected to extract deep patterns in higher dimensionality. The nature of the selected models will help in investigating the robustness and stability of the feature representations and the learning algorithms more accurately and on a larger scale.

The CNN-based architecture performs well in extracting patterns in higher dimensionality when the pattern location is irrelevant.
Therefore, we use the CNN model with image-, CFG adjacency-, and CFG algorithmic-based feature representations. On the other hand, the DNN-based architecture is used with the static-based vector representations, including Strings-, Symbols-, and Hexdump-based feature representations.

% In the following, a brief description of each learning algorithm is provided.

% \BfPara{Logistic Regression (LR)}
% LR models a binary dependent variable, known as binary classification (``0'' or ``1''), using a logistic function.
% Given ($X,Y$) as an input training set, LR trains to classify segments as positive (``1'') and negative (``0'') by estimating and optimizing the boundary between the two classes (``0'', and ``1'') and minimizing the following function:
% $$\text{\sf Loss}(f(X),Y) = \begin{cases} 
% -\log(f(X)),~Y = 1\\
% -\log(1-f(X)),~Y \neq 1
% \end{cases}, $$
% where $f(X)$ is the LR model current prediction, and $Y$ is the ground truth labels.

% \BfPara{Random Forest (RF)}
% RF learning algorithm allows for variance reduction in the output of the individual trees and mitigates the effect of noise on the training process.
% RF consists of $N$ decision trees and is used with non-linear classification tasks. Each tree is trained on random features to allow for variance reduction in the individual trees' output and decreases the effect of noise on the training process.
% The final prediction is calculated by a majority prediction vote of the decision trees or by the average prediction of all the trees.

\begin{table}
\centering
\caption{Accuracy (\%) of the baseline models. Each representation is evaluated using LR, RF, and NN-based classifiers. Note that almost all representations hold high performance (up to 99\%) in detecting IoT malware.}\vspace{-1mm}
\label{tab:baselineAccuracy}
\scalebox{0.95}{
\begin{tabular}{|l|l|c|c|c|}
\Xhline{2\arrayrulewidth}
Type & Feature              & LR     & RF   & NN                 \\
\Xhline{2\arrayrulewidth}
Binary & Image & 99.90 & 99.81  & 100\\\Xhline{2\arrayrulewidth}

CFG & Adjacency & 91.67 & 89.90  & 92.25\\\Xhline{1\arrayrulewidth}
CFG & Algorithmic & 90.20 & 99.22  & 92.09\\\Xhline{2\arrayrulewidth}
CODE & String & 98.48 & 99.43  & 98.48\\\Xhline{1\arrayrulewidth}
CODE & Symbols & 98.77 & 99.43  & 97.82\\\Xhline{1\arrayrulewidth}
CODE & Sections & 100 & 100 & 58.16\\\Xhline{1\arrayrulewidth}
CODE & Segments & 98.39 & 100  & 58.16\\\Xhline{1\arrayrulewidth}
CODE & Hexdumps & 98.96 & 99.24  & 98.48\\\Xhline{1\arrayrulewidth}
CODE & Combined & 100 & 99.90  & 57.79\\

\Xhline{2\arrayrulewidth}

\end{tabular}\vspace{-3mm}
}
\end{table}

% \BfPara{Convolutional Neural Network (CNN)}
% CNN is a powerful deep learning model used in image classification and pattern recognition. A convolution layer, which generates feature maps, is the basic unit of the CNN network. Once a feature vector is fed into a convolutional layer, it becomes abstracted to a feature map, with the shape of (feature map height) $\times$ (feature map width) $\times$ (feature map depth). 
% CNN performs well in extracting patterns in higher dimensionality when the pattern location, in the feature space, is irrelevant.
% Therefore, we use the CNN model with image-, CFG adjacency-, and CFG algorithmic-based feature representations. 

% \BfPara{Deep Neural Networks (DNN)}
% DNN model is used to extract deep encoded patterns and contains multiple consecutive fully connected layers. In the learning stage, the model configures the parameters of each single layer $l$, denoted by:
% \begin{equation}
%     h^{(l)} = a(W^{(l)}\times X + b^{(l)}),
% \end{equation}
% where, for a layer $l$, $a(.)$ is the activation function, $W^{(l)}$ is the weights of the features, and $b^{(l)}$ is the bias. We use the DNN model with the static-based representations, including Strings-, Symbols-, and Hexdumps-based representations.

\BfPara{Training Stage} The dataset is split into 80\% training and 20\% testing. The Neural Network (NN) classifiers were trained with ten epochs, and a learning rate of 0.01.

\subsection{Evaluation \& Results}
To better understand the robustness of the IoT malware detection systems, we evaluate each of the settings separately.

\subsubsection{Baseline Evaluation}
We implemented the baseline classifiers on our dataset (see~\autoref{sec:Dataset}).~\autoref{tab:baselineAccuracy} shows the performance of the classifiers. Eight out of the nine representations achieve a high detection accuracy of 99\% with at least one learning algorithm. The only exception is the CFG-based adjacency matrix representation, with an accuracy of 92.25\%. We recall that high accuracy does not reflect accurate learning, nor the quality of the learned patterns. 

\begin{figure*}[t]
    \centering
    \begin{subfigure}[t]{0.32\textwidth}
        \centering
        \includegraphics[width=0.90\textwidth]{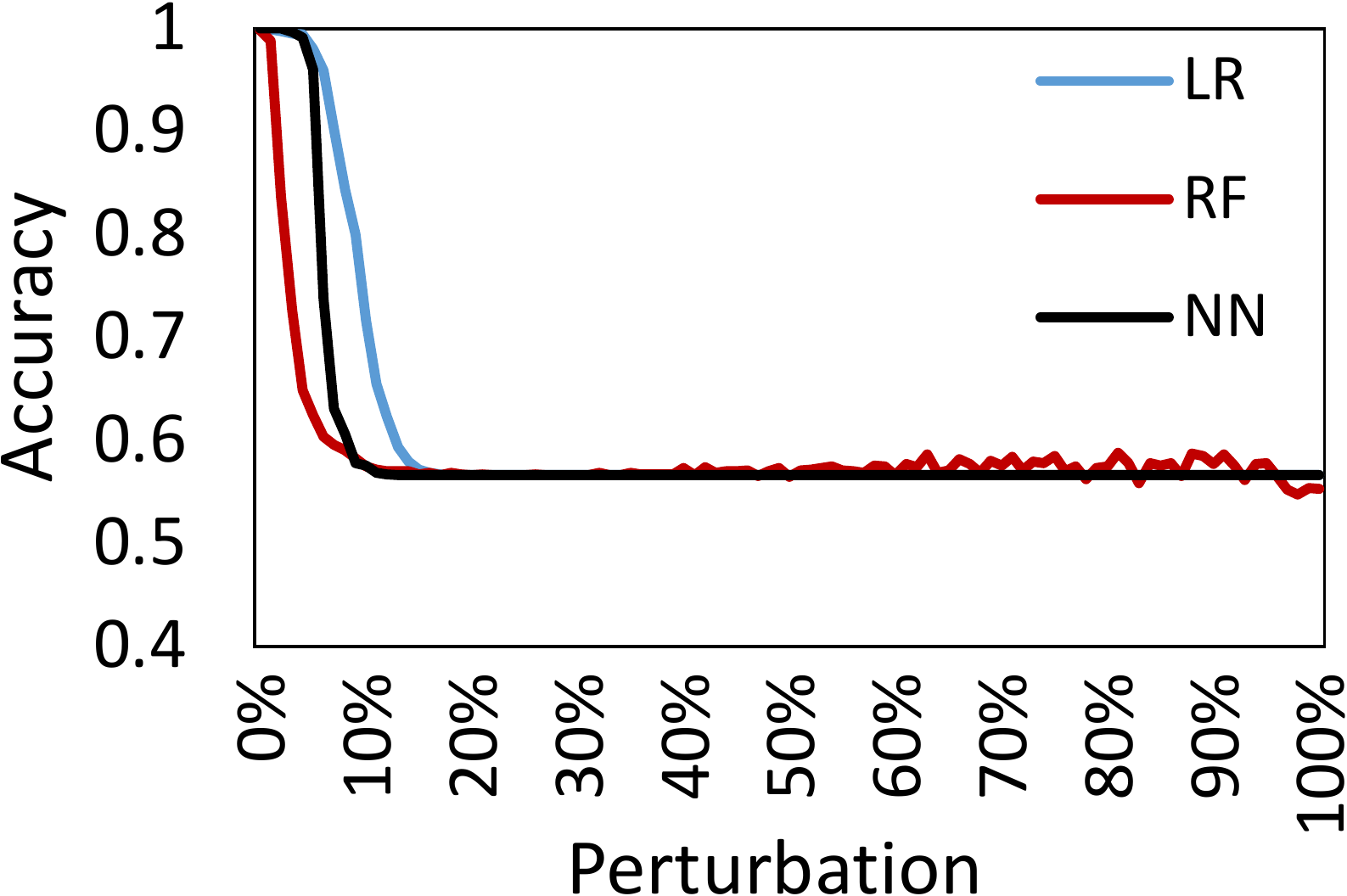}\vspace{-1mm}
        \caption{Image representation.}
        \label{fig:imageNoise}
    \end{subfigure}%
    ~ 
    \begin{subfigure}[t]{0.32\textwidth}
        \centering
        \includegraphics[width=0.90\textwidth]{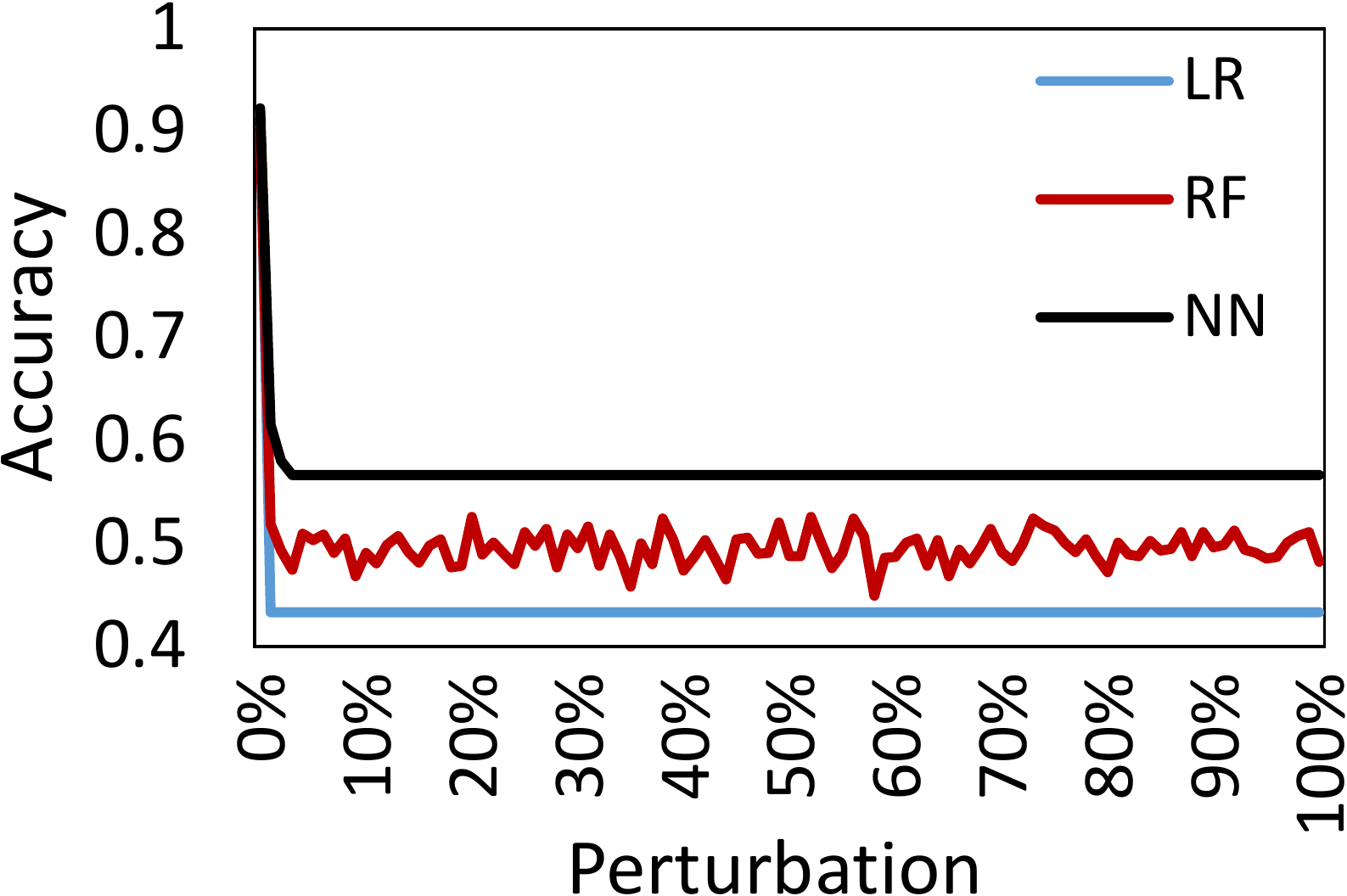}\vspace{-1mm}
        \caption{Adjacency matrix representation.}
        \label{fig:AdjacencyNoise}
    \end{subfigure}
    ~ 
    \begin{subfigure}[t]{0.32\textwidth}
        \centering
        \includegraphics[width=0.90\textwidth]{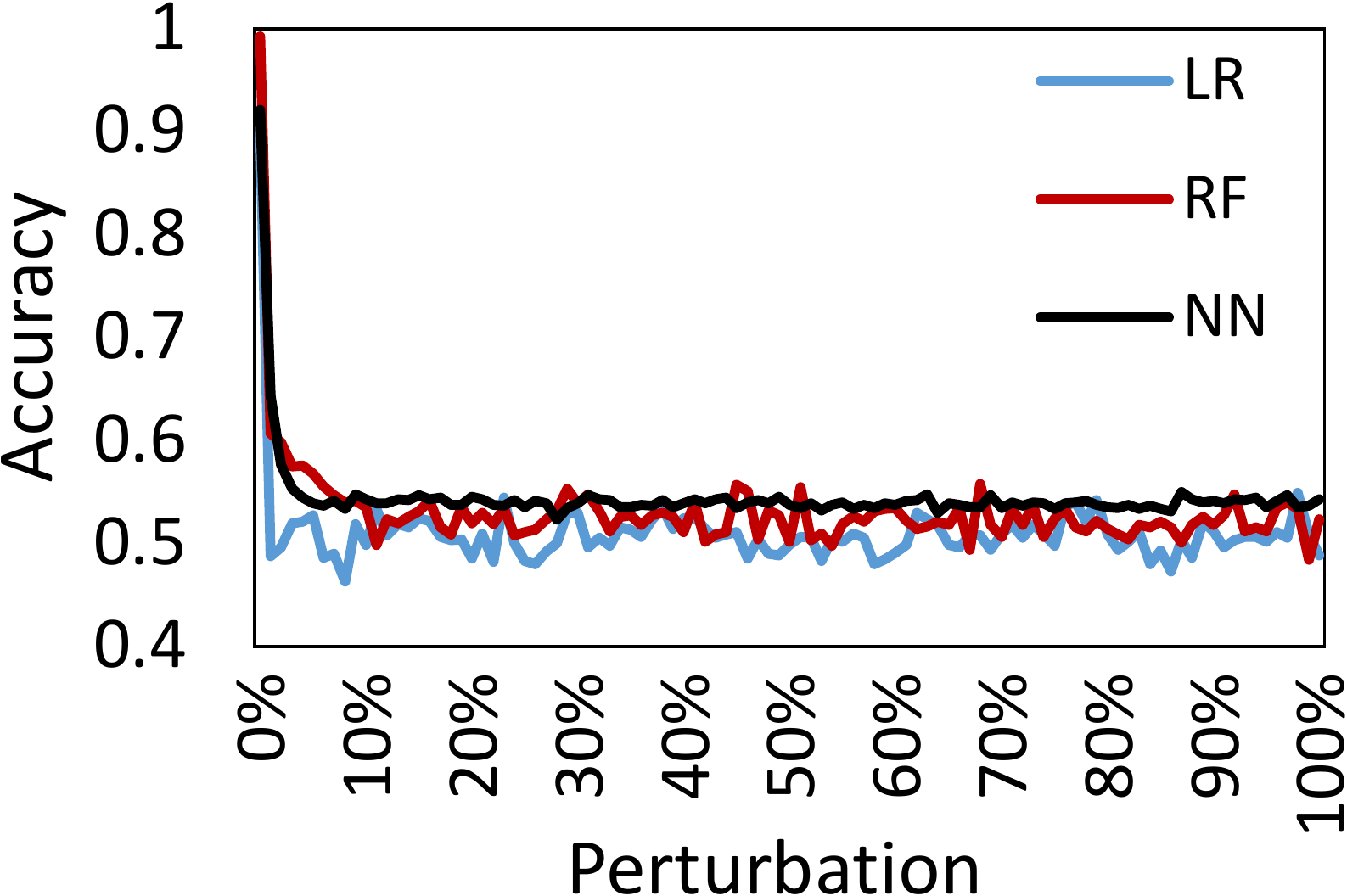}\vspace{-1mm}
        \caption{Graph algorithmic features.}
        \label{fig:AlgorithmicNoise}
    \end{subfigure}
    \begin{subfigure}[t]{0.32\textwidth}
        \centering
        \includegraphics[width=0.90\textwidth]{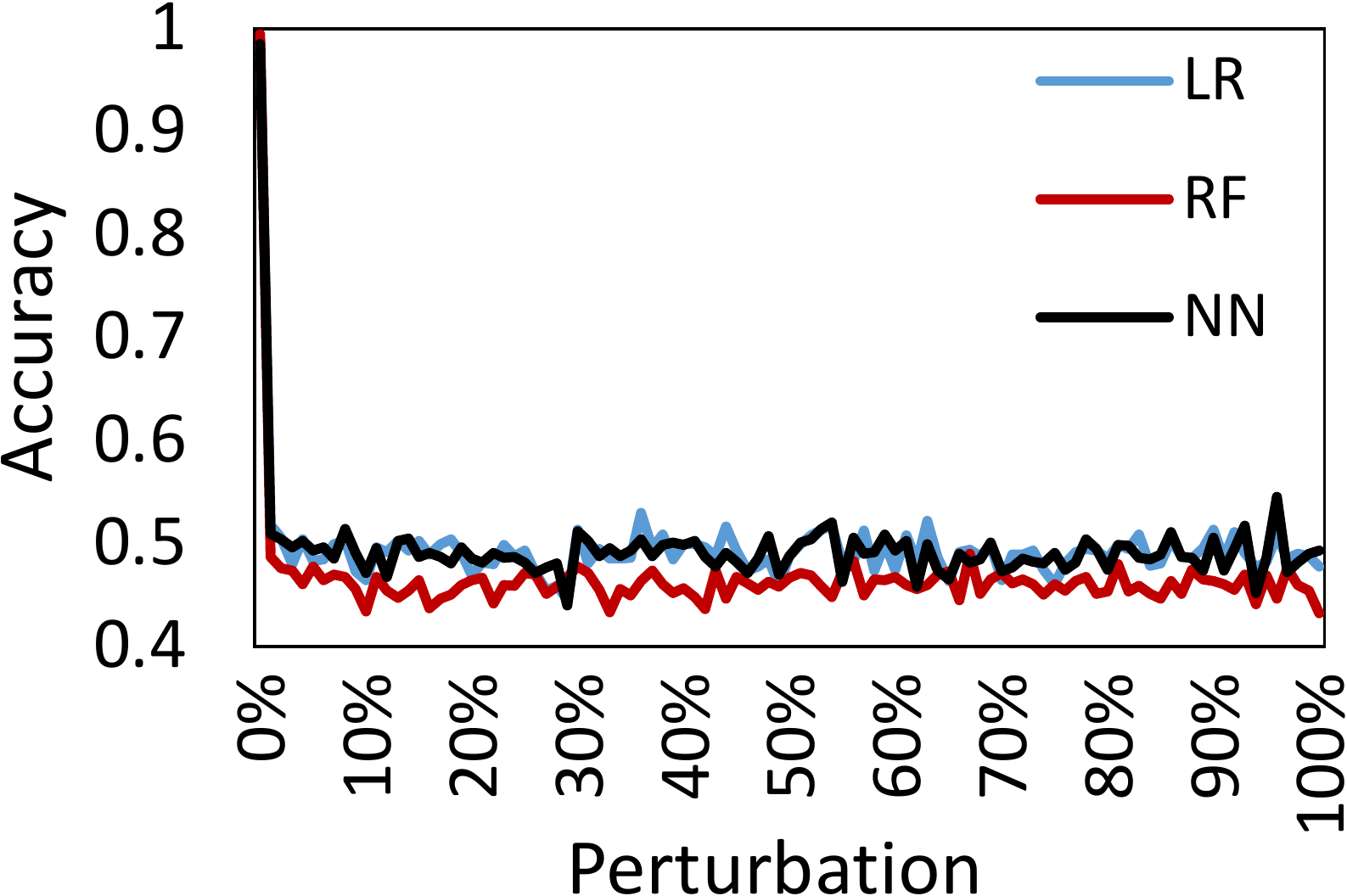}\vspace{-1mm}
        \caption{String representation.}
        \label{fig:StringNoise}
    \end{subfigure}%
    ~ 
    \begin{subfigure}[t]{0.32\textwidth}
        \centering
        \includegraphics[width=0.90\textwidth]{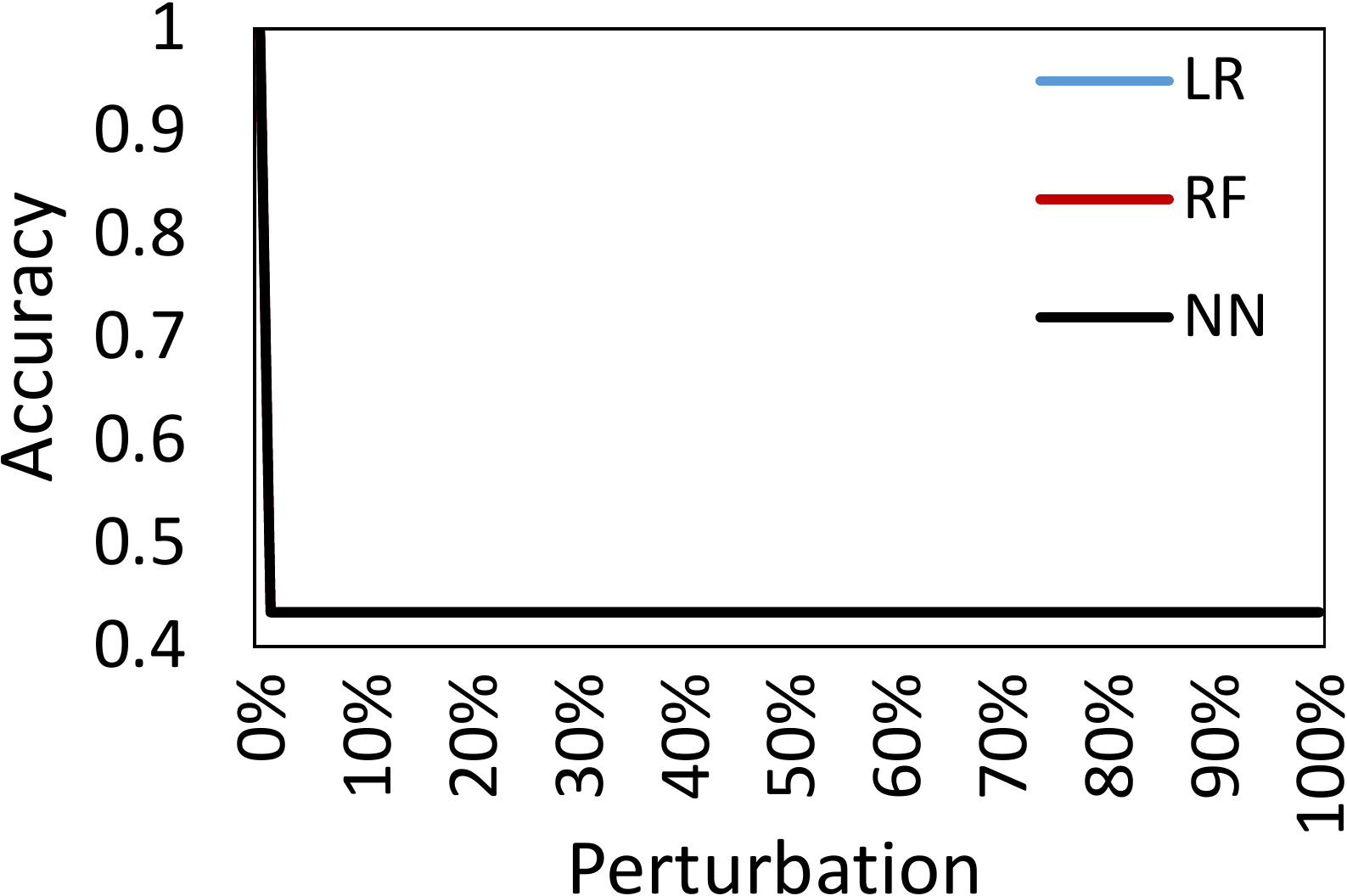}\vspace{-1mm}
        \caption{Symbols table representation.}
        \label{fig:SymbolsNoise}
    \end{subfigure}
    ~ 
    \begin{subfigure}[t]{0.32\textwidth}
        \centering
        \includegraphics[width=0.90\textwidth]{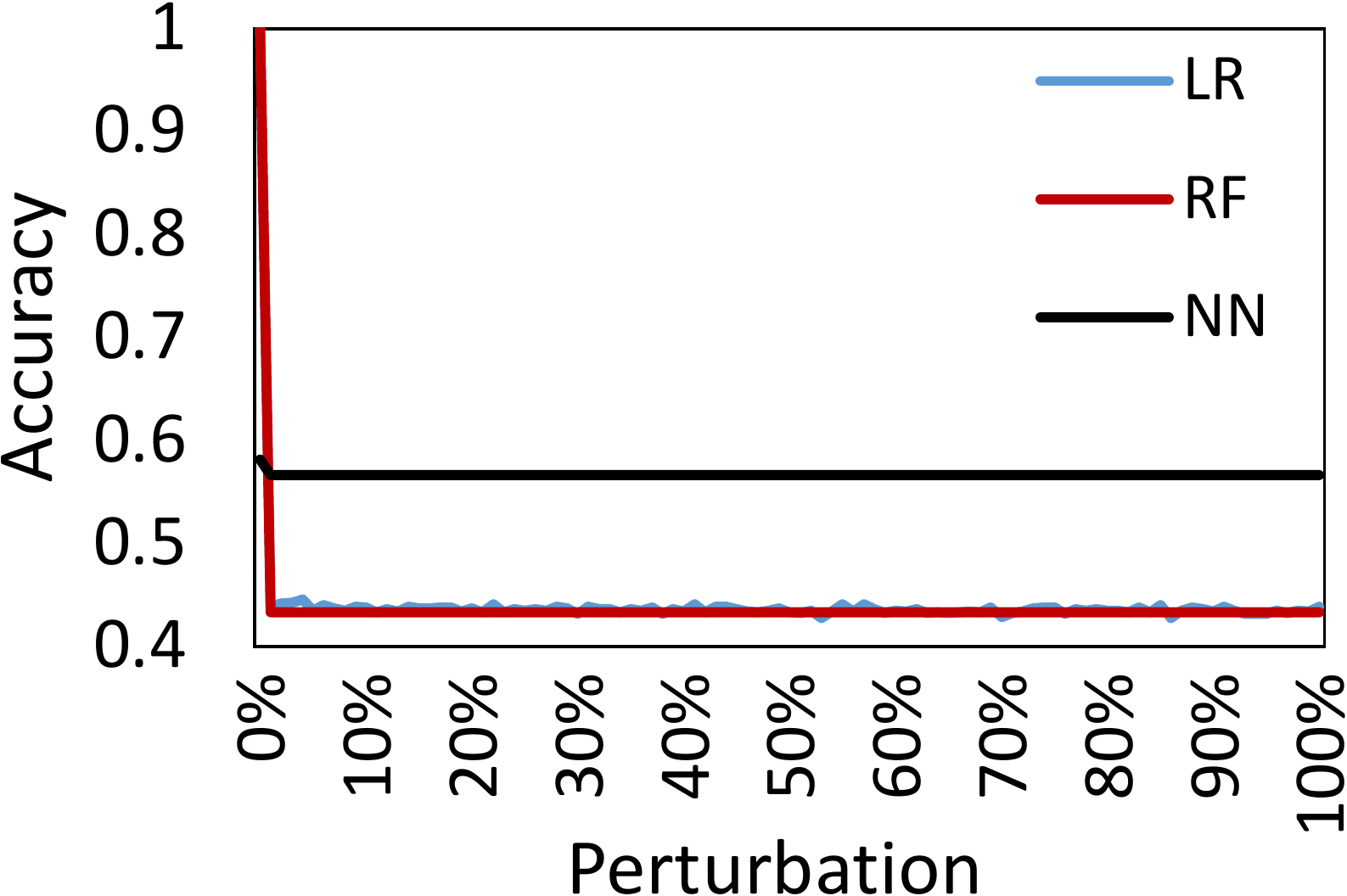}\vspace{-1mm}
        \caption{Sections table representation.}
        \label{fig:SectionsNoise}
    \end{subfigure}
    \begin{subfigure}[t]{0.32\textwidth}
        \centering
        \includegraphics[width=0.90\textwidth]{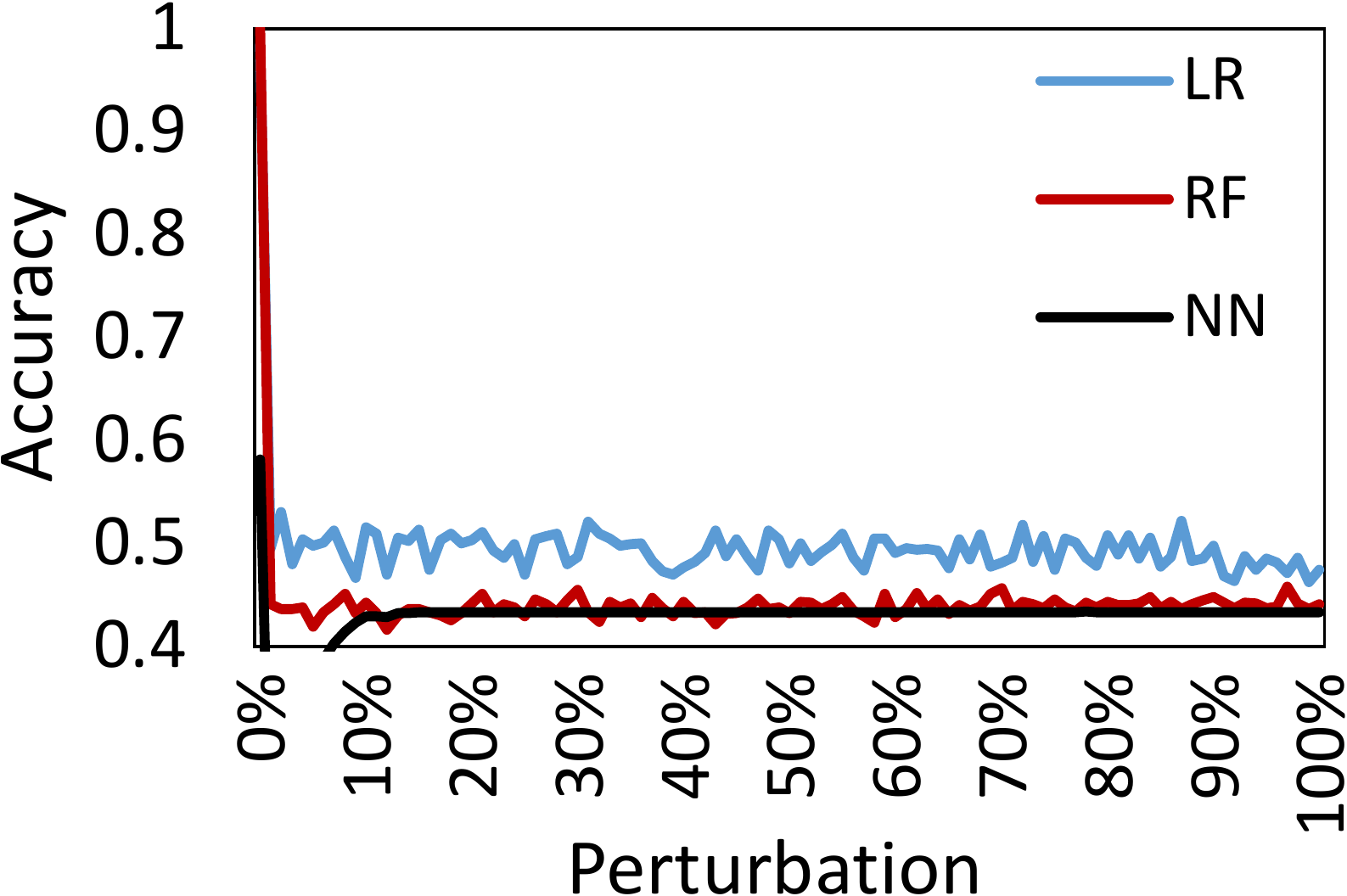}\vspace{-1mm}
        \caption{Segments table representation.}
        \label{fig:SegmentsNoise}
    \end{subfigure}%
    ~ 
    \begin{subfigure}[t]{0.32\textwidth}
        \centering
        \includegraphics[width=0.90\textwidth]{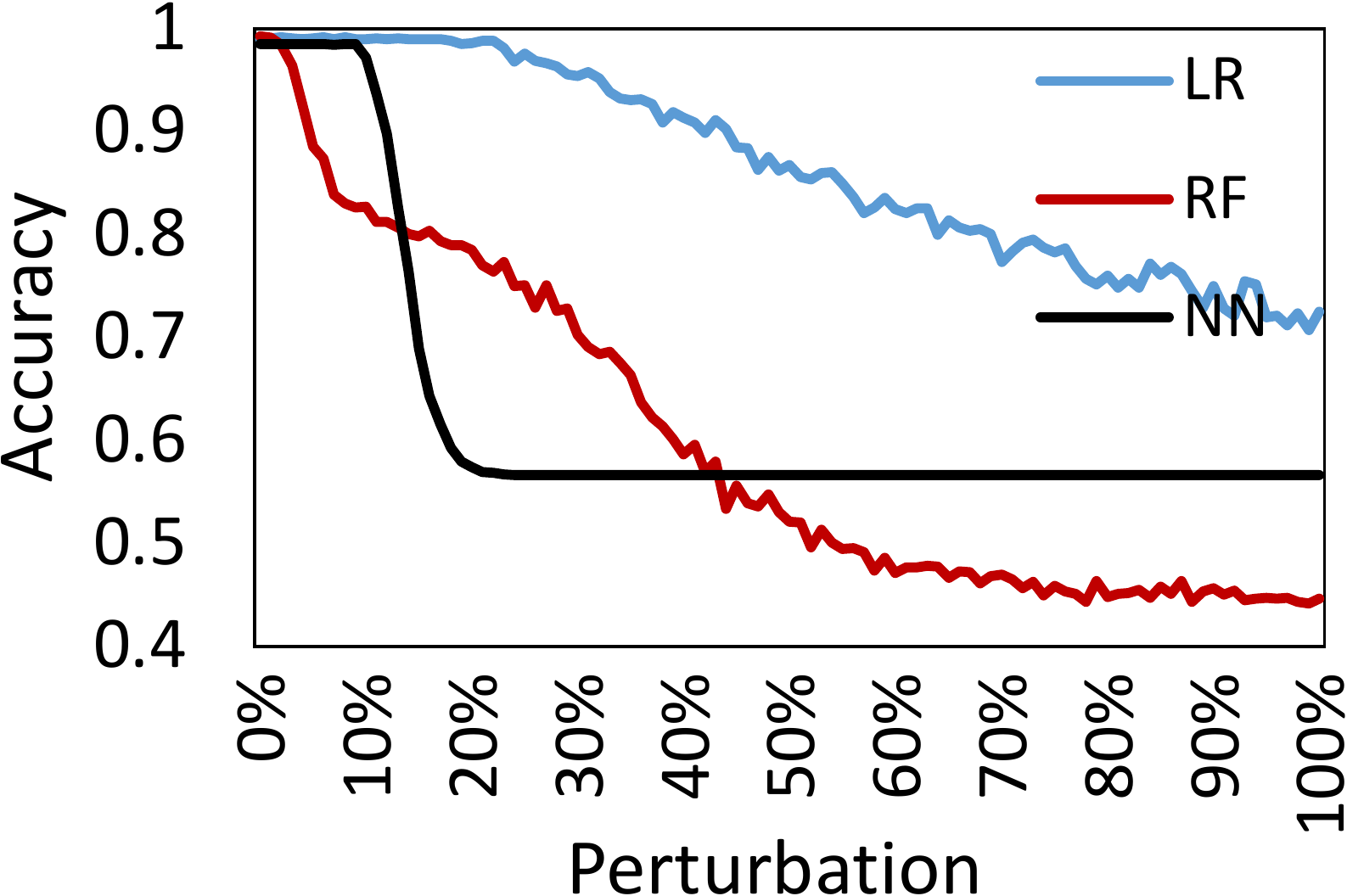}\vspace{-1mm}
        \caption{Hexdump-based representation.}
        \label{fig:HexdumpNoise}
    \end{subfigure}
    ~ 
    \begin{subfigure}[t]{0.32\textwidth}
        \centering
        \includegraphics[width=0.90\textwidth]{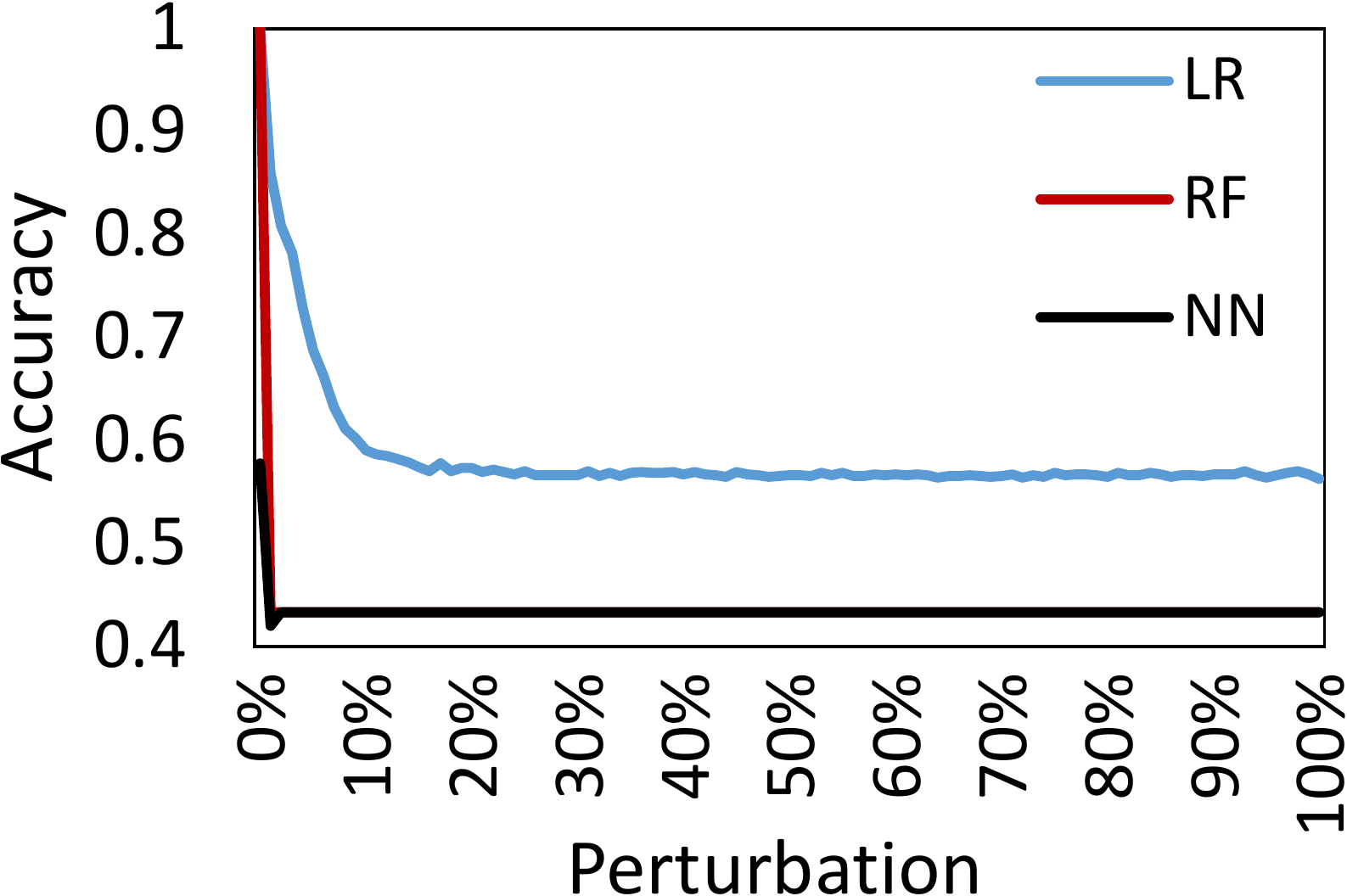}\vspace{-1mm}
        \caption{Combined static representations.}
        \label{fig:CombinedNoise}
    \end{subfigure}\vspace{-2mm}
    \caption{Baseline classifiers evaluation under various Gaussian noise perturbation rates (1\%-100\%).
    }
    \label{fig:noiseEval}\vspace{-3mm}
\end{figure*}

\subsubsection{Model Stability}
\noindent \textbf{RQ1.} \textit{``Are the baseline models correctly trained with no over-fitting and under-fitting?''}

A stable model's performance should ideally decrease linearly with the increase of the perturbation size, to eventually reach random  (50\% given the two-class classification).~\autoref{fig:noiseEval} shows the evaluation of the baseline classifiers under the Gaussian noise with 1\%-100\% perturbation. Except for the Hexdump representation, with the introduction of a perturbation size of $1\%\leq\delta\leq5\%$, the classifiers fail to deliver beyond the random guess. This highlights that the used representations are not stable and may fail due to the temporal changes in the data over time. A likely reason for this is the frequent appearance of different versions of the same or identical malware, thereby forcing the model to over-fit on the \textit{exact match} instead of extracting feasible patterns.

\vspace{1mm}
\noindent\fbox{%
    \parbox{0.478\textwidth}{%
        \textit{Key Finding:} Except for Hexdump-based representation, the baseline classifiers demonstrate high instability in their performance under small perturbation (1\% Gaussian noise).
    }%
}%\vspace{1mm}
\newline

\begin{table}[t]
\centering
\caption{Baseline classifiers evaluation under white-box settings. Only realistic and practical adversarial attacks are considered. All attacks are done on the NN and transferred to the LR- and RF-based classifiers.}\vspace{-1mm}
\label{tab:wbattacks}
\scalebox{0.85}{
\begin{tabular}{|l|l|l|l|c|}
\Xhline{2\arrayrulewidth}
Type & Feature     & Attack Type  & Model & Accuracy (\%)\\
\Xhline{2\arrayrulewidth}
\multirow{3}{*}{Binary} & \multirow{3}{*}{Image}& Transferred & LR & 63.73 \\\cline{3-5}

& & Transferred & RF & 72.71 \\\cline{3-5}

& & Direct & CNN & 63.73 \\
\Xhline{2\arrayrulewidth}
\multirow{3}{*}{CFG} & \multirow{3}{*}{Adjacency}& Transferred & LR & 81.77 \\\cline{3-5}

& &  Transferred & RF & 79.60 \\\cline{3-5}

& &  Direct & CNN & 81.30 \\
\Xhline{2\arrayrulewidth}
\multirow{3}{*}{CFG} & \multirow{3}{*}{Algorithmic}& Transferred & LR & 59.95 \\\cline{3-5}

& &  Transferred & RF & 60.70 \\\cline{3-5}

& &  Transferred & CNN & 59.95 \\
\Xhline{2\arrayrulewidth}
\multirow{3}{*}{CODE} & \multirow{3}{*}{String}& Transferred & LR & 29.08 \\\cline{3-5}

& & Transferred & RF & 30.02 \\\cline{3-5}

& & Direct & DNN & 30.59 \\

\Xhline{2\arrayrulewidth}

\end{tabular}
}\vspace{-4mm}
\end{table}

\subsubsection{White-box Attacks}

\noindent \textbf{RQ2.} \textit{``Are the classifiers prone to practical white-box adversarial attacks?''} 

Evaluating the classifiers against white-box settings is essential to understand their point-of-failure. In this context, we evaluate the white-box attacks that can be implemented directly on the binaries, or on the source code by the malware author.~\autoref{tab:wbattacks} shows the evaluation of the baseline models under white-box attacks, including binary padding and graph and string manipulation. While the binary padding can be also applied to the remaining representations (as shown later), it is considered as a white-box attack on the image-based representation only, and therefore reported here. We note that all considered attacks are implemented on the NN-based classifier, and transferred to other learning algorithms. The CFG-based algorithmic representation was evaluated using the perturbation generated on the adjacency-based representation (\ie transferred) due to their feature dependencies.

\vspace{1mm}
\noindent\fbox{%
    \parbox{0.478\textwidth}{%
        \textit{Key Finding:} For several representations, practical white-box attacks are possible, and can be transferred to related learning algorithms and representations.
    }%
}%\vspace{1mm}
\newline

\begin{table*}
\centering
\caption{Baseline evaluation under binary manipulation (\%). Packed*: optimized packing, L.A.: learning algorithm.}\vspace{-1mm}
\label{tab:RepresentationsEval}
\scalebox{0.85}{
\begin{tabular}{|l|l|l||c|c|c|c|c||c|c|c|c|c|}
\Xhline{2\arrayrulewidth}
\multirow{2}{*}{Type} & \multirow{2}{*}{Feature}   & \multirow{2}{*}{L.A.}                               & \multicolumn{5}{c||}{Benign} & \multicolumn{5}{c|}{Malware} \\\cline{4-13}
& & & Original& Packed & Packed*& Stripped& Padded&Original& Packed & Packed*& Stripped& Padded\\
\Xhline{2\arrayrulewidth}
\multirow{3}{*}{Binary} & \multirow{3}{*}{Image}  & LR & 100     & 3.92    & 4.35  & 6.31  & 63.73   & 99.83  & 98.00  & 98.00  & 98.00  & 98.33  \\\cline{3-13}
& & RF & 99.56     & 2.39    & 2.17  & 2.39  & 72.71  & 100  & 96.66  & 96.66  & 92.00  & 85.00  \\\cline{3-13}
& & NN & 100     & 6.31    & 6.31  & 2.17  & 63.73  &  100  & 100  & 100  & 100  & 100  \\\Xhline{2\arrayrulewidth}
\multirow{3}{*}{CFG} & \multirow{3}{*}{Adjacency}  & LR & 87.36     & 33.11    & 33.55  & 87.36  & 87.36  & 95.50  & 77.33  & 77.50  & 95.50  & 95.50   \\\cline{3-13}
& & RF & 88.01     & 98.91    & 99.12  & 88.01  & 88.01  & 91.50  & 73.16  & 73.16  & 91.50  & 91.50   \\\cline{3-13}
& & NN & 86.92     & 1.74    & 1.74  &  86.92  &  86.92  & 96.33  & 79.16  & 79.16  & 96.33  & 96.33  \\\Xhline{2\arrayrulewidth}

\multirow{3}{*}{CFG} & \multirow{3}{*}{Algorithmic}  & LR & 91.54     & 1.96    & 1.96  & 91.54  & 91.54  & 89.04  & 89.86  & 89.64  & 89.04  & 89.04   \\\cline{3-13}
& & RF & 99.51     & 99.56    & 99.78  & 99.51  & 99.51  & 98.96  & 88.76  & 88.76  & 98.96  & 98.96   \\\cline{3-13}
& & NN & 93.23     & 2.17    & 2.17  & 93.23  & 93.23  &  91.11  & 91.85  & 91.62  & 91.11  & 91.11  \\\Xhline{2\arrayrulewidth}

\multirow{3}{*}{CODE} & \multirow{3}{*}{String}  & LR & 96.51     & 3.48    & 3.48  & 96.51  & 96.51  & 100  & 100  & 100  & 100  & 100   \\\cline{3-13}
& & RF & 98.69     & 2.39    & 2.39  & 98.69  & 98.69  & 100  & 100  & 100  & 100  & 100   \\\cline{3-13}
& & NN & 96.51     & 0.00    & 0.00  & 96.51  & 96.51  & 100  & 100  & 100  & 100  & 100  \\\Xhline{2\arrayrulewidth}

\multirow{3}{*}{CODE} & \multirow{3}{*}{Symbols}  & LR & 97.16     & 1.08    & 1.08  & 97.16  & 97.16  & 100  & 100  & 100  & 100  & 100   \\\cline{3-13}
& & RF & 98.69     & 2.17    & 2.17  & 98.69  & 98.69  & 100  & 100  & 100  & 100  & 100   \\\cline{3-13}
& & NN & 94.98     & 3.26    & 3.26  & 94.98  & 94.98  & 100  & 100  & 100  & 100  & 100  \\\Xhline{2\arrayrulewidth}

\multirow{3}{*}{CODE} & \multirow{3}{*}{Sections}  & LR & 100     & 100    & 100  & 3.48  & 100  & 100  & 34.66  & 34.66  & 100  & 100   \\\cline{3-13}
& & RF & 100     & 3.48    & 3.48  & 100  & 100  & 100  & 100  & 100  & 100  & 100   \\\cline{3-13}
& & NN & 0.00     & 0.00    & 0.00  & 0.00  & 0.00  & 100  & 100  & 100  & 100  & 100  \\\Xhline{2\arrayrulewidth}

\multirow{3}{*}{CODE} & \multirow{3}{*}{Segments}  & LR & 96.51     & 0.00    & 0.00  & 96.51  & 96.51  & 99.83  & 99.83  & 99.83  & 99.83  & 99.83   \\\cline{3-13}
& & RF & 100     & 3.48    & 3.48  & 100  & 100  & 100  & 100  & 100  & 100  & 100   \\\cline{3-13}
& &  NN & 3.48     & 3.48    & 3.48  & 3.48  & 3.48  & 100  & 100  & 100  & 100  & 100  \\\Xhline{2\arrayrulewidth}

\multirow{3}{*}{CODE} & \multirow{3}{*}{Hexdumps}  & LR & 98.03     & 97.60    & 97.60  & 98.03  & 98.03  & 99.66  & 86.16  & 86.16  & 99.66  & 99.66   \\\cline{3-13}
& & RF & 98.25     & 1.74    & 1.74  & 98.25  & 98.25  & 100  & 92.83  & 92.83  & 100  & 100   \\\cline{3-13}
& & NN & 96.51     & 0.00    & 0.00  & 96.51  & 96.51  & 100  & 100  & 100  & 100  & 100  \\\Xhline{2\arrayrulewidth}

\multirow{3}{*}{CODE} & \multirow{3}{*}{Combined}  & LR & 100     & 3.48    & 3.48  & 3.48  & 100  & 100  & 100  & 100  & 100  & 100   \\\cline{3-13}
& & RF & 99.78     & 3.26    & 3.26  & 99.56  & 99.78  & 100  & 100  & 100  & 100  & 100   \\\cline{3-13}
& &  NN & 0.00     & 0.00    & 0.00  & 0.00  & 0.00  & 100  & 100  & 100  & 100  & 100  \\\Xhline{2\arrayrulewidth}

\end{tabular}
}\vspace{-3mm}
\end{table*}

\subsubsection{Binary Manipulation Attacks}
These settings include evaluating the classifiers under manipulation attacks on the software. 
Here, we consider binary packing under default and optimized (packing*) conditions, stripping, and padding.~\autoref{tab:RepresentationsEval} shows the evaluation results under these manipulation attacks strategies. 
In the following, we interpret these results posed as research questions.

\noindent \textbf{RQ3.} \textit{``Does binary packing affect the performance of the baseline classifiers?''}

The evaluation results show that most of the classifiers identify packed software as malicious. This indicates that they identify packing as a malicious pattern. This observation is in line with Aghakhani~\etal~\cite{aghakhani2020malware}, demonstrating that the industry-standard windows malware detection systems identify the packed software as malicious.
However, our results bring forward an exception, where Hexdump-based LR classifier maintains its performance under the two levels of packing.

\vspace{1mm}
\noindent\fbox{%
    \parbox{0.478\textwidth}{%
        \textit{Key Finding:} Baseline classifiers, in general, identify packing as malicious behavior.
    }%
}%\vspace{1mm}
\newline

\noindent \textbf{RQ4.} \textit{``Does stripping affect the baseline classifiers?''}

Recall that stripping removes information, such as the debug information, from the software binaries. However, the results exhibit that the performance of most of the representations, such as the CFG, strings, and Hexdump, are intact.

\vspace{1mm}
\noindent\fbox{%
    \parbox{0.478\textwidth}{%
        \textit{Key Finding:} Generally, existing approaches maintain high accuracy under binary stripping.

    }%
}%\vspace{1mm}
\newline

\noindent \textbf{RQ5.} \textit{``Does padding affect the  baseline classifiers?''}

Given that with binary padding we do not remove any existing functional codebase, it does not affect the analyses of the software. Therefore, it only affects the binary/image-based representation.

\vspace{1mm}
\noindent\fbox{%
    \parbox{0.478\textwidth}{%
        \textit{Key Finding:} Binary padding only reduces the performance of binary/image-based classifiers and can be countered by reverse-engineering the software samples.

    }%
}%\vspace{1mm}
\newline

\noindent \textbf{RQ6.} \textit{``Which of the representations and learning algorithms are best suited for malicious IoT software detection?''}

To answer this question, we considered the following metrics: (1) Baseline accuracy. A detector should have a minimal detection error (\ie false positive and negative rates). (2) Performance consistency. The performance of the classifiers should be robust to various binary manipulation techniques. (3) Model stability. The robustness of the classifier should encompass Gaussian noise, to some extent. Altogether, the classifier that performed best is the Hexdump-based LR classifier, followed by the CFG algorithmic-based RF classifier.

\vspace{1mm}
\noindent\fbox{%
    \parbox{0.478\textwidth}{%
        \textit{Key Finding:} Hexdump-based LR classifier is the most robust classifier, providing a stable 98.96\% baseline accuracy. 

    }%
}%\vspace{1mm}

\begin{figure*}[t]
    \centering
    \begin{subfigure}[t]{0.19\textwidth}
        \centering
        \includegraphics[width=0.99\textwidth]{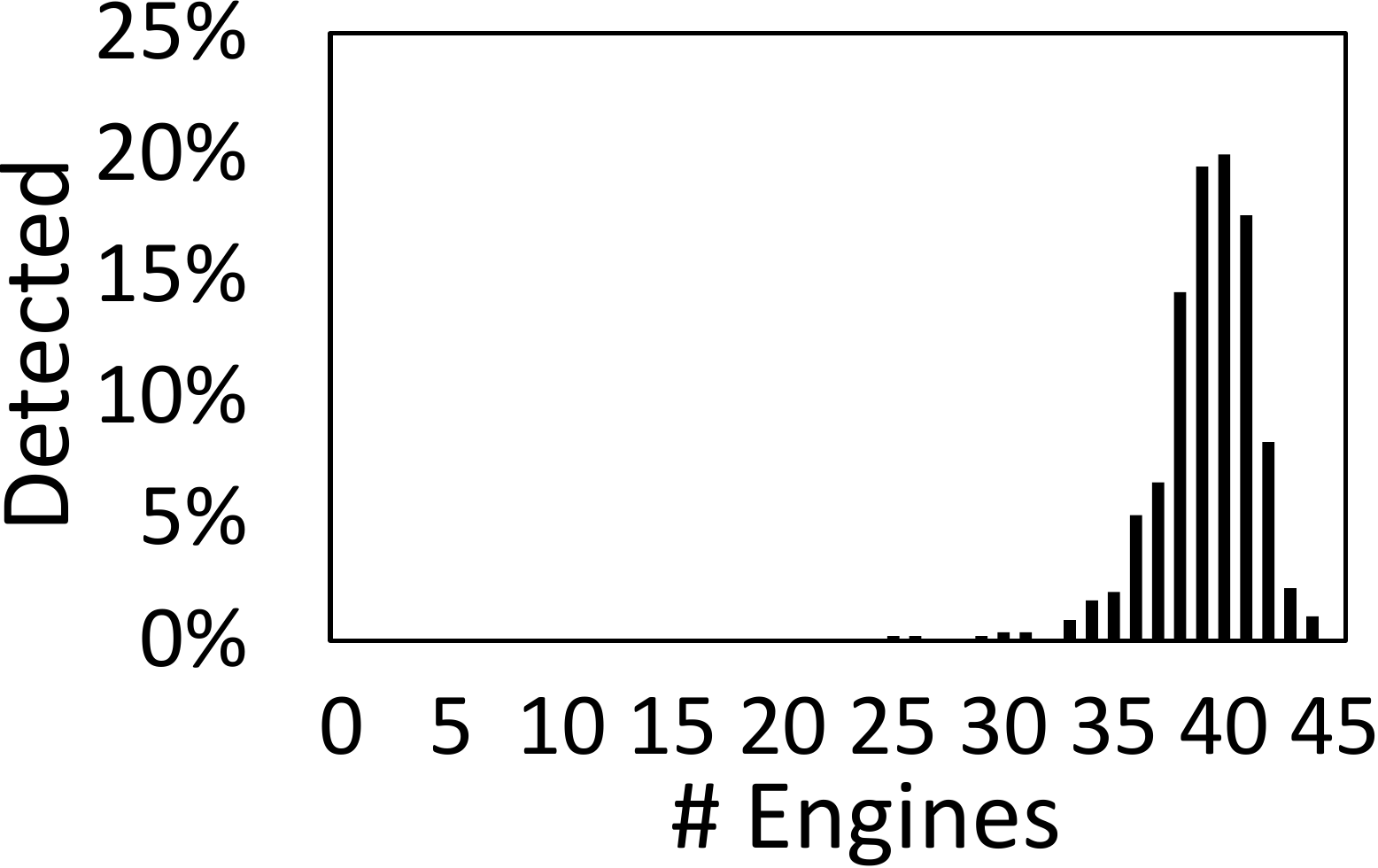}%\vspace{-1mm}
        \caption{Original.}
        \label{fig:OriginalDistribution}
    \end{subfigure}%
    \begin{subfigure}[t]{0.19\textwidth}
        \centering
        \includegraphics[width=0.99\textwidth]{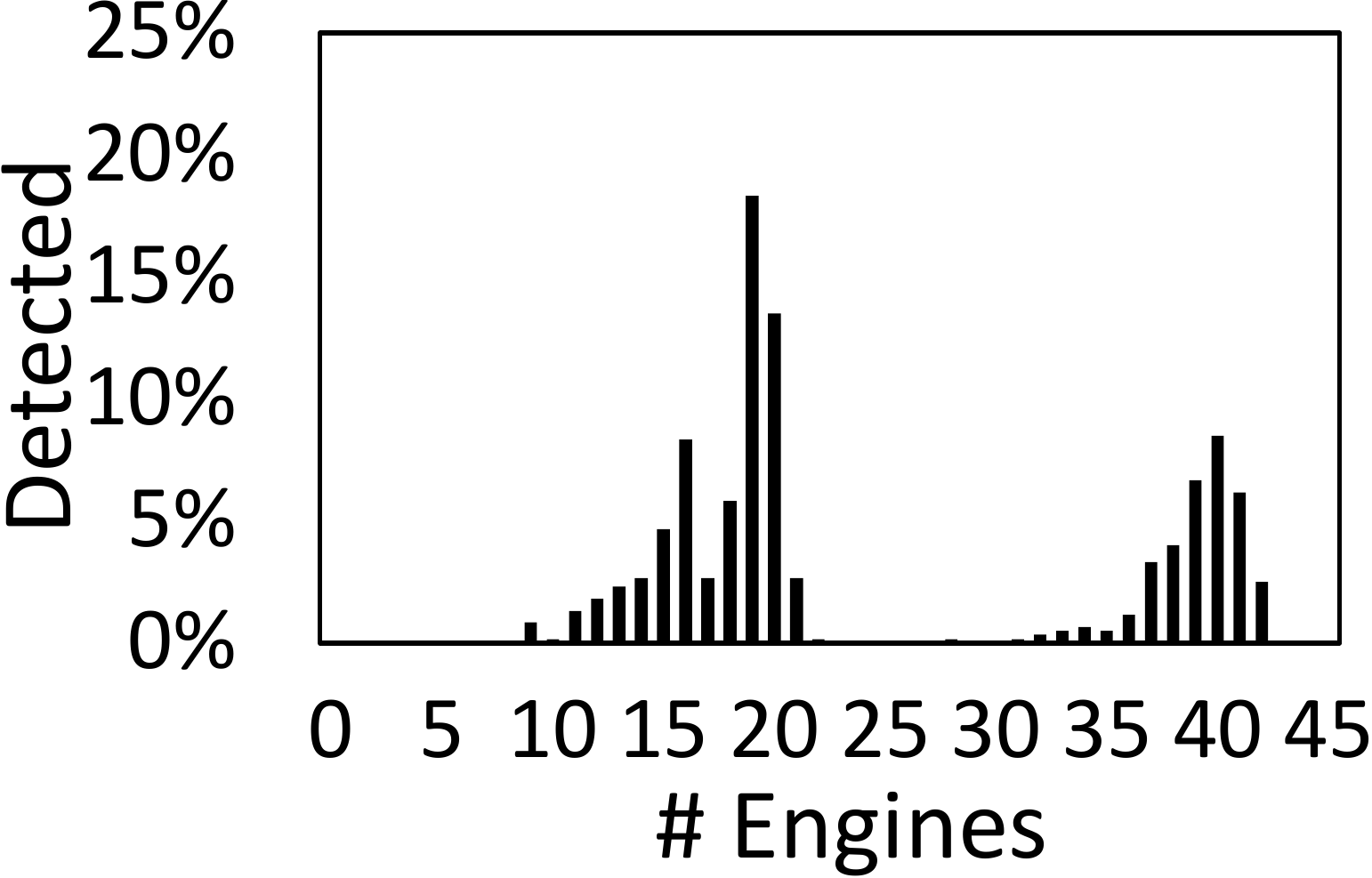}%\vspace{-1mm}
        \caption{Binary Packed.}
        \label{fig:packedMalwareDistribution}
    \end{subfigure}%
    \begin{subfigure}[t]{0.19\textwidth}
        \centering
        \includegraphics[width=0.99\textwidth]{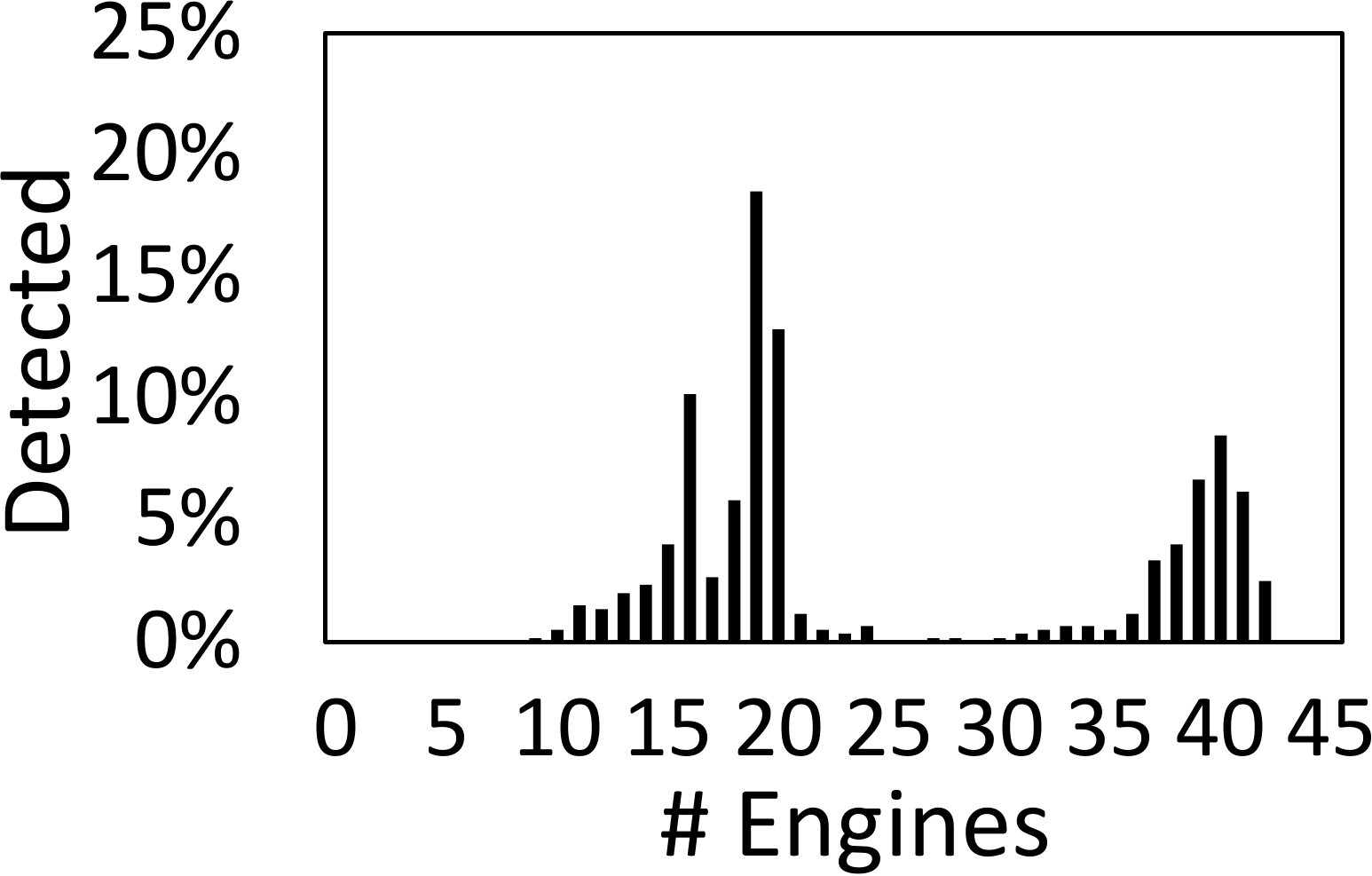}%\vspace{-1mm}
        \caption{Binary Packed*.}
        \label{fig:packedMalwareBestDistribution}
    \end{subfigure}
    \begin{subfigure}[t]{0.19\textwidth}
        \centering
        \includegraphics[width=0.99\textwidth]{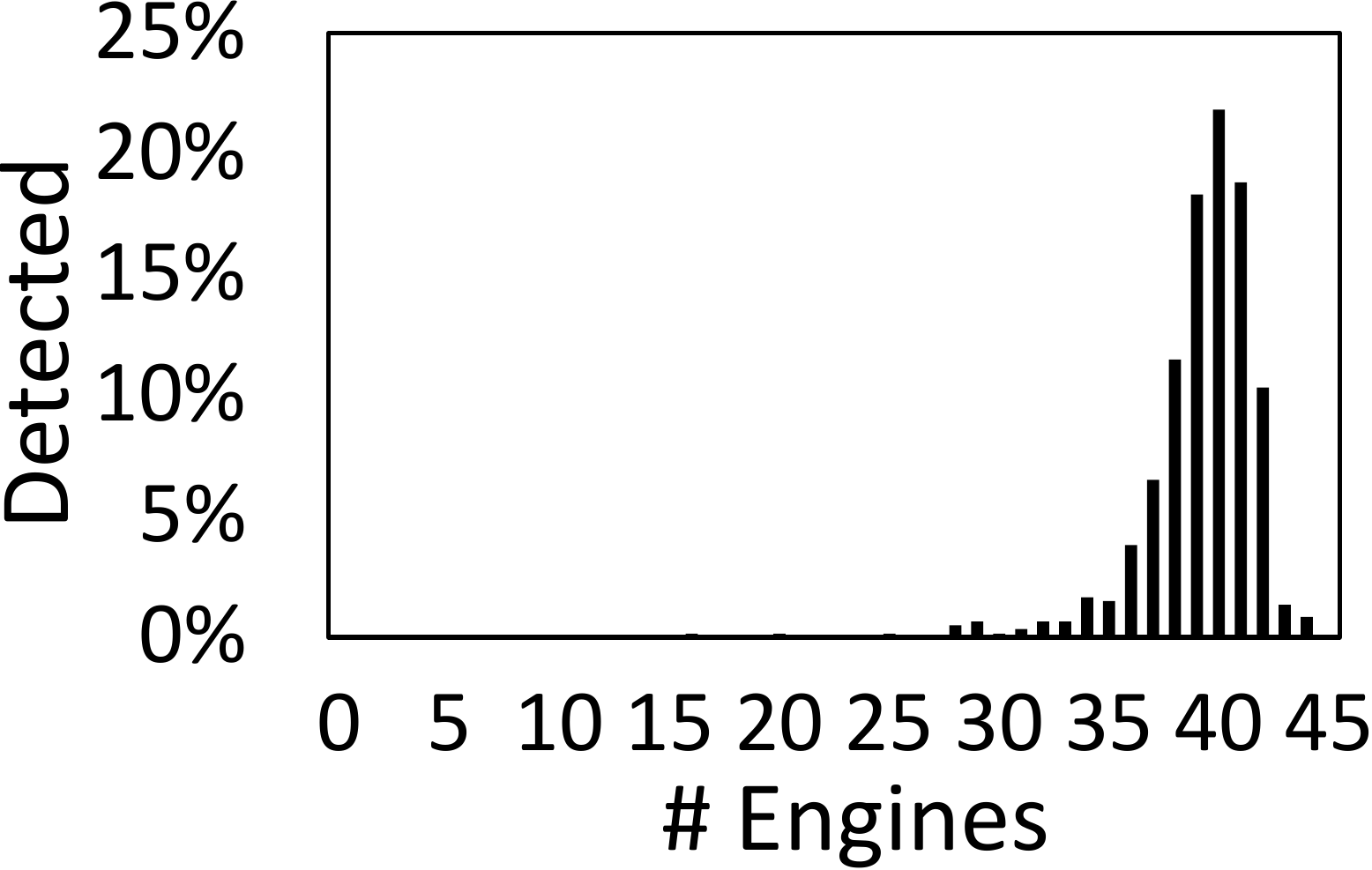}%\vspace{-1mm}
        \caption{Binary Stripped.}
        \label{fig:strippedMalwareDistribution}
    \end{subfigure}
    \begin{subfigure}[t]{0.19\textwidth}
        \centering
        \includegraphics[width=0.99\textwidth]{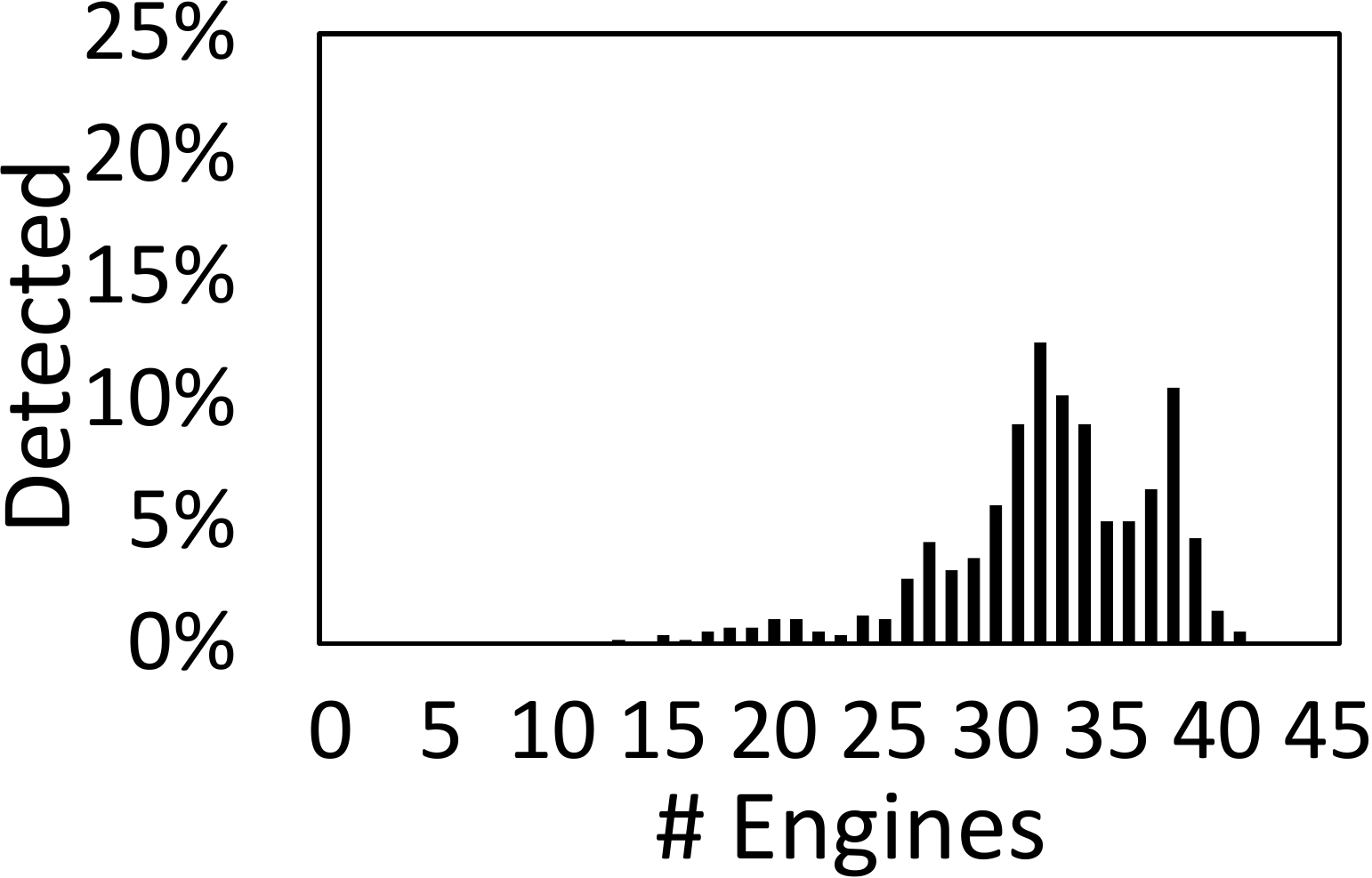}%\vspace{-1mm}
        \caption{Binary Padded.}
        \label{fig:paddedMalwareDistribution}
    \end{subfigure}\vspace{-1mm}
    \caption{The online engines' detection rate of the original and binary manipulated IoT malware samples.}
    \label{fig:DistributionOnlineEngines}\vspace{-3mm}
\end{figure*}

\section{Industry-Standard Detection Engines Robustness}\label{sec:onlineengines}
Malware authors check their software on the online detection engines to ensure that it evades the scanning engines. Given that these scan engines provide results for a pool of anti-virus engines, evading the detection from these engines is considered as a prototype for malware evolution. These mutations are then used in malware campaigns in the future. We argue that a practical malware detector should detect such mutations, or at least cover for the low-effort based mutations. 
\subsection{Experimental Setup}
Online scan engines, such as VirusTotal, are commonly used by researchers to inspect software. VirusTotal reports contain the detection results of a pool of state-of-the-art anti-virus engines that can be considered as the up-to-date capability of industry-standard malware detectors. Overall, it contains reports from 66 IoT malware detection engines. Therefore, to have a comprehensive evaluation of the existing IoT malware detectors, we also evaluate the industry-standard malware detection systems.

\BfPara{VirusTotal Reporting} 
% To evaluate the online engines, 
The original and manipulated software were uploaded to VirusTotal using their Large File Scan API. To account for the time the AI engines take to properly scan the uploaded files, we wait for 24-hours before gathering the reports. Each of the reports contains details about the uploaded file, including the date, size, header information, and the scan results of each available detection engine. Each report contains results of multiple engines (45-66), each highlighting if it detects the file as malicious or otherwise. Additionally, we found two engines that report for less than ten samples, which we removed from our list. 
Ultimately, we scan the malicious and benign software through 64 detection engines.

\BfPara{AI-based Engines} 
The next step is to separate the AI-based engines from other engines. This step is challenging as the detection engines are unlikely to share their detection approaches with the public. We manually inspect each detection engine website, searching for the used approaches. Engines that explicitly mention AI or ML are labeled as AI ($\checkmark$), while others are labeled as uncertain (\xmark).  

\BfPara{Ethical Considerations} As stated by VirusTotal, the API is not meant to be used to compare between the engines, nor be used to draw conclusions of whether engine X is better than engine Y. Toward this, we take the following considerations: (1) All engines are renamed as ``E --- $i$'', where $i$ is a given index for the engine. (2) The usage of the API is to assert that state-of-the-art scan engines are vulnerable and behave similar to the research-based detection approaches discussed in~\autoref{sec:results}. We do not intend to compare the engines, nor raise concerns against any specific service provider.

\begin{table*}
\centering
\caption{The online IoT malware detection engines evaluation (\%). Packed*: optimized packing.}\vspace{-1mm}
\label{tab:OnlineEngines}
\scalebox{0.85}{
\begin{tabular}{|l|c||c|c|c|c|c||c|c|c|c|c|}
\Xhline{2\arrayrulewidth}
\multirow{2}{*}{Engine}               & \multirow{2}{*}{AI}                    & \multicolumn{5}{c||}{Benign} & \multicolumn{5}{c|}{Malware} \\\cline{3-12}
& & Original& Packed & Packed*& Stripped& Padded&Original& Packed & Packed*& Stripped& Padded\\
\Xhline{2\arrayrulewidth}
E --- 1              & $\checkmark$ & 100             & \underline{86.41 }          & \underline{89.68 }           & 100             & 100           & 100              & 82.79            & 82.94             & 100              & 100            \\\Xhline{1\arrayrulewidth}
E --- 2            & $\checkmark$ & 100             & 100           & 100            & 100             & 100           & 98.33              & 33.83            & 34.67            & 97.33              & \underline{23.5}             \\\Xhline{1\arrayrulewidth}
E --- 3                & $\checkmark$ & 100             & 100           & 100            & 100             & 100           & 99.5               & \underline{34.67}            & \underline{35.5}              & 98.5               & \underline{37.0}             \\\Xhline{1\arrayrulewidth}
E --- 4              & $\checkmark$ & 100             & 100           & 100            & 100             & 100           & 99.33              & 94.5             & 96.33             & 99.33              & 95.29            \\\Xhline{1\arrayrulewidth}
E --- 5          & $\checkmark$ & 100             & ---               & ---                & 100             & 100           & 100              & 100            & 100             & 100              & 100            \\\Xhline{1\arrayrulewidth}
E --- 6          & $\checkmark$ & 100             & 100           & 100            & 100             & 100           & 99.67              & 99.67            & 99.67             & 99.66              & 99.67

           \\\Xhline{1\arrayrulewidth}
E --- 7              & $\checkmark$ & 100             & 100           & 100            & 100             & 100           & \underline{0.0}                & 0.0              & 0.0               & 0.0                & 0.0              \\\Xhline{2\arrayrulewidth}
E --- 22               & \xmark                   & 100             & 100           & 100            & 100             & 100           & 80.61              & 29.15            & 29.34             & 79.16              & \underline{4.04}             \\\Xhline{1\arrayrulewidth}
E --- 23           & \xmark                    & 100             & 100           & 100            & 100             & 100           & 99.67              & 99.67            & 99.5              & 99.5               & 97.33            \\\Xhline{1\arrayrulewidth}
E --- 24               & \xmark                    & 100             & 100           & 100            & 100             & 100           & 50.34              & 29.36            & 29.88             & \underline{85.21}              & 59.97            \\\Xhline{1\arrayrulewidth}
E --- 25             & \xmark                    & 100             & 100           & 100            & 100             & 100           & 84.8               & 28.42            & 28.52             & 81.27              & 4.65             \\\Xhline{1\arrayrulewidth}
E --- 26             & \xmark                   & 100             & 100           & 100            & 100             & 100           & 100              & 58.29            & 58.66             & 98.99              & 40.37            \\\Xhline{1\arrayrulewidth}
E --- 27               & \xmark                   & 100             & \underline{85.84 }          & \underline{90.07 }           & 100             & 100           & 100              & 82.78            & 82.8              & 100              & 100            \\\Xhline{1\arrayrulewidth}
E --- 28               & \xmark                    & 100             & 100           & 100            & 100             & 100           & 99.83              & 99.83            & 99.83             & 99.66              & 95.41         
   \\\Xhline{1\arrayrulewidth}

E --- 29         & \xmark                    & 100             & 100           & 100            & 100             & 100           & \underline{0.0}                & 0.0              & 0.0               & 0.0                & 0.0           
\\\Xhline{2\arrayrulewidth}
\end{tabular}
}%\vspace{0mm}
\end{table*}

\subsection{Evaluation \& Results}
\begin{figure}[t]
    \centering
    \includegraphics[width=0.450\textwidth]{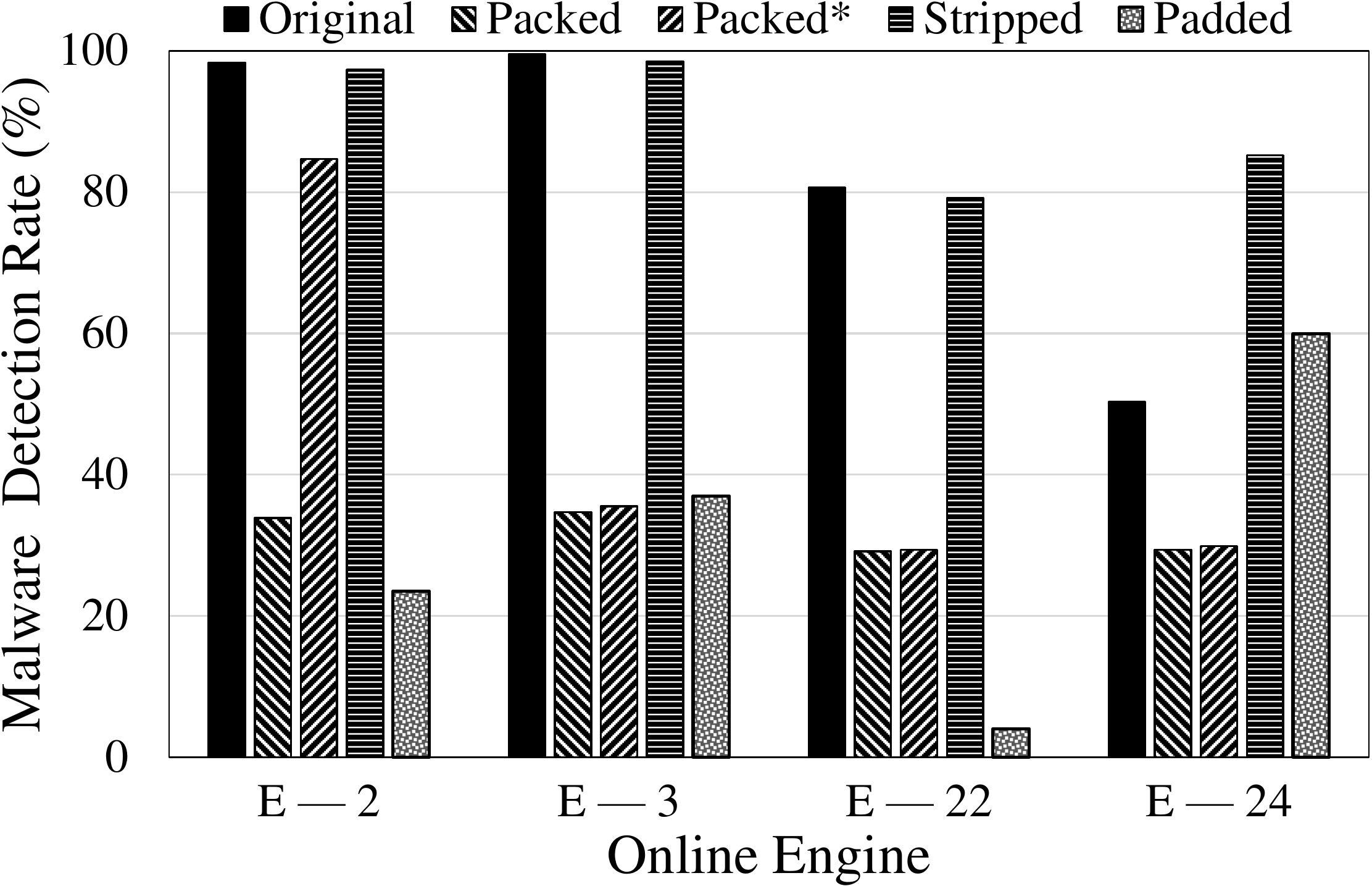}%\vspace{-3mm}
    \caption{Industry-standard detection engines robustness highlight. Binary packing significantly reduces the detection rate of Malware software (``E --- 2''). Binary stripping does not result in noticeable performance degradation, and may increase the malware detection rate (``E --- 22''). Simple binary padding to the end of the file may cause significant degradation in the performance (``E --- 3'' and ``E --- 22'').}
    \label{fig:OnlineEnginesSummary}\vspace{-4mm}
\end{figure}

In the following, we interpret the results of the industry-standard malware detectors to understand their behavior, shown in~\autoref{tab:OnlineEngines} and presented as research questions. Due to space constraint, we only report detailed results for 15 engines.
% We color-code the table for better interpretability and present analyses accordingly. 
% The detailed evaluation of the 64 engines is shown in~\autoref{tab:FullOnlineEngines} in the appendix, and 
The major insights are illustrated in~\autoref{fig:OnlineEnginesSummary}.

\noindent \textbf{RQ7.} \textit{``Does manipulation affect malware detection rate?''}

To answer this question, we recorded the number of engines that identify malware as malicious. We begin by probing the original malware samples: \autoref{fig:OriginalDistribution} shows the distribution of their detection rate by the engines. Notice that malware, on average, is detected by 40 engines, with a majority of them being detected by 35-45 engines. For the manipulated samples, however, the detection rate varies highly. \autoref{fig:DistributionOnlineEngines} shows the distribution of malicious samples by the number of engines for each of the manipulation strategies. We notice that stripping (\autoref{fig:strippedMalwareDistribution}) does not affect the distribution of the samples. However, packing (\autoref{fig:packedMalwareDistribution} and~\autoref{fig:packedMalwareBestDistribution}) highly affects the detection rate. Moreover, while binary padding had minimal effects on the baseline classifiers' performance  (\autoref{sec:results}), it highly affects their detection among the online engines. This indicates that several engines use binary-based representations (\eg binary sequence and image) to detect malicious software.

\vspace{1mm}
\noindent\fbox{%
    \parbox{0.478\textwidth}{%
        \textit{Key Finding:} Except for binary stripping, binary manipulation highly decreases the detection confidence.

    }%
}%\vspace{1mm}
\newline

\noindent \textbf{RQ8.} \textit{``How individual engines generally perform?''}

To answer this question, we evaluate each individual detection engine using the original and manipulated benign and malicious software, shown in~\autoref{tab:OnlineEngines}. We observe that multiple engines perform poorly, with 36\% of the engines (23 out of 64) failing in identifying malware ($\approx$ 0\% accuracy), such as ``E --- 7'' and ``E --- 29''.
Additionally, except for ``E --- 1'' and ``E --- 27'', the benign detection accuracy is 100\%, similar trends were observed for packed, stripped, and padded benign software.

\vspace{1mm}
\noindent\fbox{%
    \parbox{0.478\textwidth}{%
        \textit{Key Finding:} Several engines (36\%) exhibit reduced performance for detecting original and binary manipulated malicious software.
    }%
}%\vspace{1mm}
\newline

\noindent \textbf{RQ9.} \textit{``Does packing affect the engines' performance?''}

The evaluations exhibit that packing does not affect the performance of the engines in accurately detecting benign software (except for ``E --- 1'' and ``E --- 27''). 
This observation is in contrast to previous observations~\cite{aghakhani2020malware} (refer to~\autoref{sec:results}).
However, packing, generally, reduces the accuracy of malware being detected as malware. For instance, ``E --- 3'' performance declined from 99.5\% to $\approx$ 35\% when tested with packed malware.
We also observed that optimized packing does not decrease the detection rate, in fact, it slightly increases the chance of malicious software being detected, as compared to the standard packing. 
Additionally, for engines, such as ``E --- 5'', we observe that no results were reported for benign packed binaries, while achieving 100\% in other categories. This can be attributed to the low confidence of the engine in labeling benign packed samples.

\vspace{1mm}
\noindent\fbox{%
    \parbox{0.478\textwidth}{%
        \textit{Key Finding:} Although packing reduces the detection rate of malicious software, it has no effect on the benign software detection rate. Optimized packing has a higher detection rate in comparison with default packing.
    }%
}%\vspace{1mm}
\newline

\noindent \textbf{RQ10.} \textit{``Does stripping affect the engines' performance?''}

% Recall that stripping misclassifies the benign software under multiple baseline representations (\autoref{sec:results}).
There is no noticeable decrease ($<$1\%) in the detection accuracy of stripped software in the case of online engines. In fact, for some engines (\ie ``E --- 24''), the malware detection performance increased from 50.34\% to 85.21\% after stripping. 

\vspace{1mm}
\noindent\fbox{%
    \parbox{0.478\textwidth}{%
        \textit{Key Finding:} Stripping has no negative effect on the performance of the engines, albeit increasing the accuracy in some instances.
    }%
}\vspace{2mm}
\newline

\noindent \textbf{RQ11.} \textit{``Does padding affect the engines' performance?''}

Binary padding significantly decreases the performance of several online engines, such as ``E --- 2'', ``E --- 3'', and ``E --- 22''. This is maybe attributed to the fact that appending binaries disrupt the existing signatures. The online engines' reports show that $>53\%$ of them are affected negatively, with $>14\%$ of them exhibiting a drastic decrease in performance ($>70\%$ decrease). 
Although padding does not affect the reverse-engineered features, the decrease in performance, regardless, indicates that the engines use the raw binary representations (\eg binary sequence- and image-based) for classification, which apparently can be easily disrupted. 

\vspace{2mm}
\noindent\fbox{%
    \parbox{0.478\textwidth}{%
        \textit{Key Finding:} Binary padding highly reduces the performance of several engines, while leaving others intact.
    }%
}\vspace{3mm}

\section{Concluding Remarks} \label{sec:conclusion}
Malware analysis and detection have been the focus of the research community and the industry alike, with many advances in defenses with the use of AI-backed systems. Despite those advances, these systems have been shown to be vulnerable to several simple-yet-effective adversarial attacks, such as binary stripping and packing. With this work, we systematically evaluate the state of a range of malware detectors, proposed by the research community and industry-standard.

Our efforts show that malware detectors proposed in the literature are vulnerable to adversarial perturbation and binary manipulation attacks. Similarly, industry-standard malware detectors are prone to such attacks. Our efforts also unveil the status-quo of the existing detectors, and bring forward various insights to consider when proposing detection systems. Particularly, in addition to optimizing baseline malware detection accuracy, researchers should take into account the robustness of the proposed systems under adversarial capabilities. This obligates for a deep understanding of the underlying learning algorithms and data representations, alongside the learned patterns and their characteristics.

\bibliographystyle{IEEEtran}

\bibliography{ref}

\end{document}